\documentclass[12pt,a4paper]{article}
\pdfoutput=1

\usepackage{jcappub}
\allowdisplaybreaks

\usepackage[margin=1in]{geometry}
\usepackage{graphicx}
\usepackage{amsfonts,amsmath}
\usepackage{bm}
\usepackage{subfigure}
\usepackage{float}
\usepackage{hyperref}
\usepackage[dvipsnames]{xcolor}
\usepackage{amsmath}
\usepackage[normalem]{ulem}
\usepackage{empheq}
\usepackage{amssymb}
\usepackage{tikz}
\usepackage{tabularx}
\newcommand{\class}{{\sc class}}

\usepackage{booktabs}

\newcommand{\tj}[6]{ \begin{pmatrix}
  #1 & #2 & #3 \\
  #4 & #5 & #6
\end{pmatrix}}
\definecolor{redViolet}{rgb}{0.8,0.0,0.4}
 \usepackage{color}

\def\lsim{~\rlap{$<$}{\lower 1.0ex\hbox{$\sim$}}}
\def\bsim{~\rlap{$>$}{\lower 1.0ex\hbox{$\sim$}}}

\def\ln{{\rm ln}}

\def\be{\begin{equation}}
\def\ee{\end{equation}}
\def\bea{\begin{eqnarray}}
\def\eea{\end{eqnarray}}
\def\ba{\begin{align}}
\def\ea{\end{align}}
\def\bi{\begin{itemize}}
\def\ei{\end{itemize}}

\newcommand{\Gaunt}[6]{{\cal G}^{#1#2#3}_{#4#5#6}}

\newcommand{\bn}{{\bm n}}

\def\Om{\Omega}

\def\De{\Delta}

\def\bx{{\bm{x}}}
\def\bk{{\bm{k}}}

\def\ba{{\vec{g}}}

\newcommand{\HH}{\mathcal{H} }

\newcommand{\cd}{\cdot}

\title{The Full-Sky Angular Bispectrum in Redshift Space}

\author[a,b]{{Enea~Di~Dio},}
\author[c]{Ruth~Durrer,}
\author[d,e]{Roy~Maartens,}
\author[f,g]{Francesco~Montanari,}
\author[e]{Obinna~Umeh}

\affiliation[a]{Lawrence Berkeley National Laboratory, 1 Cyclotron Road, Berkeley, CA 93720, USA}
\affiliation[b]{Berkeley Center for Cosmological Physics, University of California, Berkeley, CA 94720, USA}
\affiliation[c]{University of Geneva, Department of Theoretical Physics and Center for Astroparticle Physics (CAP), 24 quai E. Ansermet, CH-1211 Geneva 4, Switzerland}
\affiliation[d]{Department of Physics $\&$ Astronomy, University of the Western Cape, Cape Town 7535, South Africa}
\affiliation[e]{Institute of Cosmology and Gravitation, University of Portsmouth, Portsmouth, PO1 3FX, United Kingdom}
\affiliation[f]{Instituto de F\'isica Te\'orica IFT-UAM/CSIC,
  Universidad Aut\'onoma de Madrid, Cantoblanco 28049 Madrid, Spain}
\affiliation[g]{University of Helsinki, Department of Physics
and Helsinki Institute of Physics
P.O. Box 64, FIN-00014 University of Helsinki, Finland}

\abstract{We compute the redshift-dependent angular bispectrum of
  galaxy number counts at tree-level, including nonlinear clustering
  bias and estimating numerically for the first time the effect of
  redshift space distortions (RSD). We show that for narrow redshift
  bins the amplitude of nonlinear RSD is comparable with the matter
  density perturbations. While our numerical results only include
    terms relevant on sub-horizon scales, the formalism can readily be
    extended to the full tree-level bispectrum. Our approach does not
  rely on the flat-sky approximation and it can be easily generalized
  to different sources by including the appropriate bias expansion.
  We test the accuracy of Limber approximation for different z-bins.
  We highlight the subtle but relevant differences in the angular
  bispectrum of galaxy number counts with respect to CMB, due to the
  different scale dependence of perturbations. Our formalism can also
  be directly applied to the angular HI intensity mapping
  bispectrum.\footnote{\label{fn:byspectrum} We release the {\sc
      Byspectrum} code (version 0.1) required to reproduce the results
    of this paper at \url{https://gitlab.com/montanari/byspectrum}.}}

\begin{document}
\maketitle

%%%%%%%%%%%%%%%%%%%%%%%%%%%%%%%%%%%%%%%%%%%%%%%%%
\section{Introduction}

The next generation of galaxy surveys will open a new era of precision cosmology and yield new capacities to study the observable Universe via  three-dimensional matter distribution. The optical/infrared surveys that are planned for Euclid~\cite{Laureijs:2011gra}, LSST~\cite{Abell:2009aa} and DESI~\cite{Aghamousa:2016zmz}, will cover huge volumes of the Universe with very high galaxy numbers, and errors  on standard cosmological parameters will be dominated by systematics. The new radio surveys planned for the SKA~\cite{Bacon:2018} will cover even larger volumes, but new challenges from systematics in the radio will need to be overcome.

These upcoming surveys are based on tremendous advances in experimental precision. In order to fully exploit their great potential, it is necessary to develop also new theoretical tools and improve theoretical precision.
In this spirit, we develop a new analysis of the angular bispectrum for  galaxy number counts in terms of directly observable quantities: redshift $z$ and angular position $\bn$.
In addition to the power spectrum, the bispectrum will be increasingly
important for improving constraints and breaking degeneracies (see
e.g.~\cite{Gil-Marin:2016wya}).  So far, most analyses have used the Cartesian
Fourier-space bispectrum (see e.g. Refs.~\cite{Jolicoeur:2017nyt,Yankelevich:2018uaz}) or the 3-point correlation function (for a recent treatment including  redshift space distortions (RSD) see e.g.~\cite{Slepian:2017lpm}), which imposes a plane-parallel
approximation, and is unable to incorporate the effects of lensing
magnification. For current surveys, this may be a reasonable
approximation. But next-generation surveys require the inclusion of
wide-angle correlations, given their large sky area, and of lensing
magnification effects, given their high redshift reach. The direct way
to include full-sky and lensing magnification effects is to use the
angular bispectrum, as explained in detail in~\cite{DiDio:2015bua}
(see
also~\cite{Kehagias:2015tda,Assassi:2017lea,Bertacca:2017dzm}).

In the analysis developed in Ref.~\cite{DiDio:2015bua} the nonlinear
clustering bias and the full effect of RSD were not included. Here we
include the nonlinear bias, and for the first time, we include the
full RSD in the numerical computation of the angular bispectrum
(up to terms only relevant on scales close to the Hubble horizon
  \cite{DiDio:2014lka} that are nonetheless straightforward, although
  cumbersome, to further incorporate in our formalism). Including
efficiently RSD perturbations in terms of the 3-point function or
angular bispectrum is a non-trivial problem, and some different
approaches have been recently proposed, see
Ref.~\cite{Slepian:2018vds}.

We derive the bispectrum of galaxy number counts in terms of directly observable
quantities -- angles and redshifts.  Our approach presents two
advantages: first, by adopting only observable quantities we do not
need to assume any cosmological model to convert angles and redshifts
into distances; second, it naturally provides a full-sky description
of the bispectrum. In particular it includes the description of effects
imprinted at the largest scales, e.g.~primordial non-Gaussianity and
relativistic lightcone effects  as well as wide-angle effects, which are typically of the same
magnitude as these other effects (see e.g.~\cite{Bertacca:2012tp,
  Gaztanaga:2015jrs, Hall:2016bmm, Lepori:2017twd, Tansella:2017rpi}).

 We also apply our formalism to describe the angular HI intensity mapping bispectrum. Indeed while HI intensity mapping foregrounds can limit the ability to perform tomography, the need for very thin redshift bins requires one to include RSD perturbations in the angular power spectrum and bispectrum of intensity mapping. At the level of bispectrum, this has never been considered previously and our approach introduces a proper description of RSD on very narrow redshift bins.

 In section~\ref{sec:gnc} we introduce the dominant terms
   contributing to galaxy number counts up to second order in redshift
   space, including galaxy bias. In section~\ref{sec:angbisp} we
   compute the angular bispectrum and present a novel derivation of
   the RSD term, well-suited for numerical estimation. In
   section~\ref{sec:numerical} we present numerical results and
   discuss the accuracy of the Limber approximation. In
   section~\ref{sec:hi_im} we explain how our formalism can be readily
   adapted for intensity mapping studies and we discuss its limitations.
   We conclude in
   section~\ref{sec:conclusions}. Appendix~\ref{app:rsd} discusses
   details about the RSD bispectrum relevant for numerical
   computations. In appendix~\ref{sec:b_dens_app} we provide an
   alternative derivation of the density bispectrum, closer to the
   novel approach introduced here for RSD. In
   appendix~\ref{sec:gener-spectra-geom} we list geometrical factors
   relevant for the angular bispectrum. Finally, in
   appendix~\ref{sec:cv} we compute cosmic variance of the galaxy number count bispectrum at lowest order.

%%%%%%%%%%%%%%%%%%%%%%%%%%%%%%%%%%%%%%%%%%%%%%%%%
\section{Galaxy number counts}
\label{sec:gnc}

A spectroscopic galaxy survey provides the redshift $z$ and the angular position $\bn$ for each source. From the number of galaxies $N \left( \bn , z \right)$ we can define the galaxy number count fluctuation as
\be
\Delta\left( \bn , z \right) = \frac{N\left( \bn , z \right) - \langle N\rangle \left( z \right)}{\langle N\rangle \left( z \right)}
\ee
where $\langle (\cdots) \rangle$ denotes the angular average at fixed observed redshift $z$.
Assuming Gaussian initial conditions, non-Gaussian correlations are
generated by non-linear gravitational clustering. To determine the
tree-level bispectrum we need to compute the fluctuations up to second
order in perturbation theory. This has been done recently in terms of
galaxy number counts
by~\cite{Yoo:2014sfa,Bertacca:2014dra,DiDio:2014lka} (see also
\cite{Fry:1999yv, Verde:2000xj} for early works).

In our work we consider a perturbed FLRW metric described by
\be
ds^2 = a^2 \left[ -\left( 1+ 2 \Psi\right) d\eta^2 + \left( 1- 2\Phi \right) d\bx^2 \right]\,,
\ee
where the purely scalar metric perturbations agree (to first order) with the gauge-invariant Bardeen potentials.
At linear order, considering only the dominant terms, which scale as $\left( k/\HH\right)^2 \Phi^{(1)} \sim \delta^{(1)}$, i.e.~the terms which dominate on sub-Hubble scales, we
have\footnote{See~\cite{Bonvin:2011bg,Challinor:2011bk} for the full relativistic expression.}
\be
\Delta^{(1)}=
b_1 \delta^{(1)} +  \HH^{-1} \partial^2_r  v^{(1)}  \,,
\ee
where $\HH= \dot a/a$ is the comoving Hubble parameter, $r( z)$ is the
comoving distance to redshift $z$, $\delta^{(1)}$ is the first-order matter density perturbation in the comoving gauge and $v^{(1)}$
  is the first-order  velocity perturbation in the longitudinal gauge.

We neglect here the  lensing term which is parametrically of the same order but usually, for $z<2$, with equal redshifts and thin z-bins, the lensing term is much smaller than density and redshift space distortion. Also, the novelty of the present work focuses on RSD.
The lensing contributions to the tree-level bispectrum have been already computed in Ref.~\cite{DiDio:2015bua}. It is worth pointing out that lensing and RSD dominate in opposite regimes. Indeed while lensing dominates the cosmological signal for large radial separation, or at high redshifts in a signal averaged over a wide redshift binning, RSD decay quickly for sources well separated in redshift and in wide redshift bins.

At second order, we use the convention $\Delta=\Delta^{(1)}+\Delta^{(2)}$. Again including only the dominant terms, which scale as $\left( k/\HH\right)^4 [\Phi^{(1)}]^2$, we have
\bea
\Delta^{(2)}(\bn,z) &=& b_1 \delta^{(2)} + \frac{1}{2}b_2
\big(\delta^{(1)}\big)^2 + b_s\, s^2 + \HH^{-1} \partial^2_r v^{(2)}
\nonumber \\
&& + \HH^{-2} \left[\left( \partial^2_r
    v^{(1)}\right)^2+ \partial_rv^{(1)}\, \partial_r^3v^{(1)} \right] + \HH^{-1}
\left[\partial_r v^{(1)}\, \partial_r \delta^{(1)}+\partial^2_r v^{(1)} \, \delta^{(1)}\right] \, , \quad
\label{2nd_counts}
\eea
where $s$ is related
  to the clustering bias tidal field (see below).
The full second-order expression is much more cumbersome,
covering several pages
(see~\cite{Yoo:2014sfa,Bertacca:2014dra,DiDio:2014lka} for number
counts, and~\cite{Umeh_inprog} for intensity mapping).  In this
`quasi-Newtonian' approximation, we can set $\Phi=\Psi$.

For the nonlinear clustering bias, we assume a local bias model
and neglect stochastic bias terms. Following
~\cite{Desjacques:2016bnm}, we use the convention
\be
\delta_g = b_1\delta+{1\over2}b_2\, \delta^2 + b_s\, s^2\,.
\ee
Here $\delta=\delta^{(1)}+\delta^{(2)}$ is the matter over-density in comoving gauge and the bias coefficients are scale-independent.
The tidal bias coefficient is $b_s$, where
 $s^2=s_{ij} s^{ij}$ and the tidal field is given by
\be
s_{ij} = \frac{2}{3 \Omega_M \HH^2} \partial_i \partial_j \Phi - \frac{1}{3} \delta_{ij} \delta \, .
\ee
In the simplest local bias model, there is no tidal bias at the time of galaxy formation (see e.g.~Ref.~\cite{Baldauf:2012hs,Desjacques:2016bnm}) which leads to
\be \label{bs}
b_s=-{2\over7}(b_1-1)\,.
\ee
\begin{figure}[t]
\begin{center}
\includegraphics[width=0.7\textwidth]{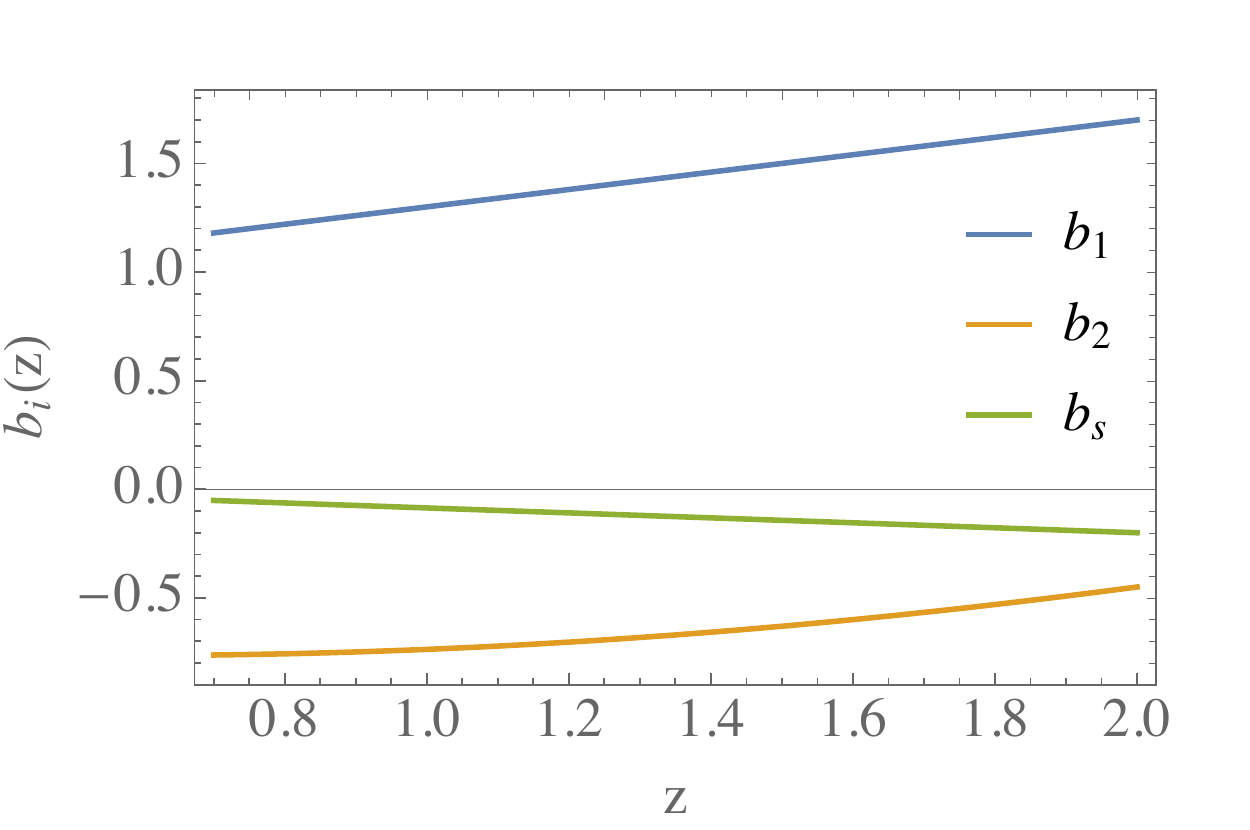}
\caption{\label{f:bias} Clustering bias coefficients for a Euclid-like survey, following Table 1 of Ref.~\cite{Yankelevich:2018uaz}, for $b_1,b_2$ and $b_s$ as a function of redshift.}
\end{center}
\end{figure}
We adopt the bias parameters presented in Ref.~\cite{Yankelevich:2018uaz} for a Euclid-like survey, converted to our bias convention.  The different bias values are shown in Fig.~\ref{f:bias}. Here we provide also a simple polynomial fit
 \bea
 b_1 \left( z\right)&=& 0.9 + 0.4 z \, ,
 \\
 b_2\left(z\right)&=& -0.704172 -0.207993 z +0.183023 z^{2}-0.00771288 z^3 \, .
 \eea

%%%%%%%%%%%%%%%%%%%%%%%%%%%%%%%%%%%%%%%%%%%%%%%%%
\section{The angular bispectrum}
\label{sec:angbisp}

In this section we compute the tree-level bispectrum induced by the 3-point function containing one second order term,\footnote{In principle we are interested in the connected part of the bispectrum, which means we have to replace $\Delta^{(2)}(\bn)$ by $\De^{(2)}(\bn)- \langle \De^{(2)}(\bn)\rangle$. But due to statistical isotropy $ \langle \De^{(2)}(\bn)\rangle$ only contributes to the  monopole and  not to the bispectrum $B_{\ell_1\ell_2\ell_3}^{m_1m_2m_3}$ for $\ell_i\neq 0$. For this reason we ignore this subtlety here.}
\bea
B(\bn_1,\bn_2,\bn_3,z_1,z_2,z_3) &\equiv& \langle \Delta \left( \bn_1, z_1 \right) \Delta \left( \bn_2, z_2 \right) \Delta \left( \bn_3, z_3 \right) \rangle\\
B_{\rm{tree}}(\bn_1,\bn_2,\bn_3,z_1,z_2,z_3) &\equiv& \langle \Delta^{(2)} \left( \bn_1, z_1 \right) \Delta^{(1)} \left( \bn_2, z_2 \right) \Delta^{(1)} \left( \bn_3, z_3 \right) \rangle +\circlearrowleft  \label{e:Btree}
\eea
where $\circlearrowleft$ denotes the two additional permutations where
the second order term is evaluated at $\left( \bn_2, z_2 \right)$ and
$\left( \bn_3, z_3 \right)$, respectively.
We expand $\De$ in spherical harmonics,
$$\De(\bn,z)=\sum_{\ell=0}^{\infty}\sum_{m=-\ell}^{\ell}a_{\ell m}(z)Y_{\ell m}(\bn)\,, \qquad a_{\ell m}(z)=\int d\Om_{\bn} \De(\bn,z)Y^*_{\ell m}(\bn) $$
so that
\be
B(\bn_1,\bn_2,\bn_3,z_1,z_2,z_3) =\sum_{\substack{\ell_1, \ell_2, \ell_3 \\ m_1, m_2,m_3} }  B^{m_1m_2m_3}_{\ell_1\ell_2\ell_3}(z_1,z_2,z_3)Y_{\ell_1m_1}(\bn_1)Y_{\ell_2m_2}(\bn_2)Y_{\ell_3m_3}(\bn_3)  \,,  \label{e:bispecamp}
\ee
where
\bea
B^{m_1m_2m_3}_{\ell_1\ell_2\ell_3}(z_1,z_2,z_3) &=& \left\langle a_{\ell_1 m_1}(z_1)a_{\ell_2 m_2}(z_2)a_{\ell_3 m_3}(z_3)\right\rangle \nonumber \\
&~\hspace*{-5.5cm}=&~\hspace*{-2.5cm}\!\!\!\! \!\int \!d\Omega_1 d\Omega_2 d\Omega_3 B \left( \bn_1, \bn_2, \bn_3, z_1, z_2 ,z_3 \right) Y^*_{\ell_1 m_1} \!\left( \bn_1 \right) Y^*_{\ell_2 m_2 } \!\left( \bn_2 \right) Y^*_{\ell_3 m_3 }\! \left( \bn_3 \right) .
\eea
Statistical isotropy demands that $B(\bn_1,\bn_2,\bn_3,z_1,z_2,z_3)$ only depends on the scalar pro\-ducts $\bn_1\cd\bn_2\,,~\bn_1\cd\bn_3$ and $\bn_1\cd\bn_2$. This dictates the $m_i$ dependence of the bispectrum,
\be\label{bi_harm_red}
B_{\ell_1\ell_2\ell_3}^{m_1m_2m_3}(z_1,z_2,z_3) =\Gaunt{m_1,}   {m_2,}   {m_3} {\ell_1,} {\ell_2,} {\ell_3} b_{\ell_1 \ell_2 \ell_3}(z_1,z_2,z_3)\,,
\ee
where $\Gaunt{m_1,}   {m_2,}   {m_3} {\ell_1,} {\ell_2,} {\ell_3} $ is the Gaunt integral  which is simply related to the Clebsch--Gordan coefficients~\cite{Abram}
\bea
\Gaunt{m_1,}   {m_2,}   {m_3} {\ell_1,} {\ell_2,} {\ell_3} &=& \int d \Omega \ Y_{\ell_1 m_1 } \left( \bn \right)Y_{\ell_2 m_2 } \left( \bn \right) Y_{\ell_3 m_3 } \left( \bn \right)  \nonumber \\
&=&\left(\begin{array}{ccc} \ell_1 & \ell_2 & \ell_3 \\ 0 & 0 & 0\end{array}\right)\left(\begin{array}{ccc} \ell_1 & \ell_2 & \ell_3 \\ m_1 & m_2 & m_3\end{array}\right)
\sqrt{\frac{(2\ell_1+1)(2\ell_2+1)(2\ell_3+1)}{4\pi}} \,.
\eea
In the last equation we have expressed the Gaunt integral in terms of the Wigner $3j$ symbols,  see e.g.~\cite{Abram}. The Gaunt integral is invariant under permutations of its columns and it is non-vanishing only if $m_1+m_2+m_3=0$ and the triangle inequality is satisfied, i.e., $|\ell_2-\ell_3|\leq \ell_1\leq \ell_2+\ell_3$. Furthermore, the sum $\ell_1+\ell_2+\ell_3$ has to be even. The quantity $b_{\ell_1\ell_2\ell_3}$ is called the reduced bispectrum. It contains all non-trivial physical information.

We assume Gaussian initial conditions so that linearly evolved perturbations are Gaussian and $\langle \De^{(1)}(\bn_1,z_1)\De^{(1)}(\bn_2,z_2)\De^{(1)}(\bn_3,z_3)\rangle =0$.  The first non-vanishing contribution to the bispectrum is therefore $B_{\rm{tree}}$ defined in (\ref{e:Btree}), and we   want to compute this contribution to $b_{\ell_1\ell_2\ell_3}(z_1,z_2,z_3)$.

Below we show the explicit form for all the different contributions to
$ \Delta^{(2)}( \bn, z)$.  In order to somewhat simplify the notation,
we derive the bispectrum first by setting
$\Delta^{(1)} \left( \bn, z \right)\sim b_1 \delta^{(1)} \left( \bn, z
\right)$. Including the linear redshift space distortion is however
simple and we do it at the end, see
eq.~\eqref{include_RSD}.

%%%%%%%%%%%%%
\subsection{RSD $\HH^{-1} \partial_r^2v^{(2)}$}
\label{sec:b_rsd}
We start considering the contribution of the pure second order redshift-space distortion. At this point we want to stress that the RSD contribution to the bispectrum has not been computed previously in terms of directly observable quantities like the full-sky angular bispectrum. Our approach does not rely on any (geometrical) approximation, like flat-sky or Limber.
Let us point out that we will consider the Newtonian kernels~\cite{Goroff:1986ep,Bernardeau:2001qr}  in the rest of the paper, which is in line with neglecting terms of higher order in $\HH/k$, see~\cite{Tram:2016cpy}. Our results can be generalized to an arbitrary (separable) non-linear velocity kernel $G_2\left( \bk_1 , \bk_2 \right)$, including the full higher order GR dynamics~\cite{Villa:2015ppa,Tram:2016cpy,Jolicoeur:2017eyi,Castiblanco:2018qsd}.

We follow the approach developed in Ref.~\cite{Assassi:2017lea} in order to exchange the order of integration,
providing a fully analytical expression for the bispectrum.
We begin with the following 3-point function
\be \label{eq:rsd_corr}
b_1(z_2) b_1(z_3) \langle \HH^{-1} \partial_r^2v^{(2)} \left( \bn_1,
  z_1 \right) \delta^{(1)} \left( \bn_2, z_2 \right) \delta^{(1)}
\left( \bn_3, z_3 \right)\rangle \,.
\ee
In Fourier space the second order velocity potential is given
by~\cite{Bernardeau:2001qr}
\be\label{e:v2}
v^{(2)} \left( \bk_1 ,z \right) =- \frac{\HH(z) f(z)}{\left( 2 \pi \right)^3k_1^2} \int d^3k_2 d^3k_3 \  \delta_D \left( \bk_1 - \bk_2 - \bk_3 \right) G_2 \left( \bk_2 , \bk_3 \right) \delta \left( \bk_2 , z \right)
\delta \left( \bk_3 , z \right) \, ,
\ee
where $f=d\ln D_+/d\ln a$ in terms of the linear growth factor $D_+$ and
scale factor $a$,\footnote{As discussed in section
  \ref{sec:numerical}, for numerical computations we interface our
  code to the \class{} Boltzmann solver, where the growth factor
  enters only implicitly in perturbation equations via the velocity
  divergence
  $\Theta^{(1)}(\bk, z)= -\HH(z) f(\bk, z) \delta^{(1)}(\bk, z)$
  \cite{Blas:2011rf}. The presence of radiation and
  neutrinos actually leads to a
  scale-dependent growth function, but we neglect this small effect. The
  second-order solutions presented here are themselves valid for an
  Einstein-de Sitter Universe, and in a $\Lambda$CDM cosmology,
  the growth factor can be simply replaced by the corresponding one, see e.g.~\cite{Bernardeau:2001qr} for details.} and \be
G_2 \left( \bk_2, \bk_3 \right) = \frac{3}{7} + \frac{1}{2}
\frac{\bk_2 \cdot \bk_3 }{k_2 k_3} \left( \frac{k_2}{k_3} +
  \frac{k_3}{k_2} \right) + \frac{4}{7} \left( \frac{\bk_2 \cdot
    \bk_3}{k_2 k_3} \right)^2 \,.\ee
    We replace the angular
dependence $\bk_2 \cdot \bk_3$ with $k^2_1=\left(\bk_2+\bk_3 \right)^2=k_2^2+k_3^3 +2\bk_2\cd\bk_3$, \bea
G_2\left( k_1 , k_2 ,k_3 \right) &=&\frac{3}{7} + \frac{1}{4} \frac{k_1^2
  - k_2^2 - k_3^2}{k_2 k_3} \left( \frac{k_2}{k_3} + \frac{k_3}{k_2}
\right) + \frac{1}{7} \left( \frac{k_1^2 - k_2^2 - k_3^2}{k_2 k_3}
\right)^2
\nonumber \\
&=& -\frac{3 \left(k_2^2-k_3^2\right)^2}{28 k_2^2 k_3^2} -
\frac{1}{28} k_1^2 \left(\frac{1}{k_2^2}+\frac{1}{k_3^2}\right) +
\frac{k_1^4}{7 k_2^2 k_3^2}
\nonumber \\
&=& G_2^{(0)} \left( k_2 , k_3 \right) + G_2^{(2)} \left( k_2 , k_3
\right) k_1^2+ G_2^{(4)} \left( k_2 , k_3 \right) k_1^4 \, .  \eea
In this way we have separated the velocity kernel $G_2$ into terms
proportional to different powers of $k_1$ which contain the angular
dependence induced by the scalar product $\bk_2 \cdot\bk_3$.

As in Ref.~\cite{DiDio:2015bua}, we now rewrite the Dirac-delta distribution in eq.~(\ref{e:v2}) as
\bea
 \delta_D \left( \bk_1 +  \bk_2 + \bk_3 \right) &=& \frac{1}{(2\pi)^3}\int d^3x e^{i( \bk_1 +  \bk_2 + \bk_3)\bx} \nonumber\\
 &=&8 \sum_{\ell'_i , m'_i} i^{\ell'_1 + \ell'_2 +\ell'_3}( -1)^{\ell'_1 + \ell'_2 + \ell'_3 } \Gaunt{m'_1,}   {m'_2,}   {m'_3} {\ell'_1,} {\ell'_2,} {\ell'_3} Y_{\ell'_1 m'_1} ( \hat \bk_1) Y_{\ell'_2 m'_2} ( \hat \bk_2) Y_{\ell'_3 m'_3} ( \hat \bk_3 )
 \nonumber \\
 &&
 \times \int_0^\infty d\chi \chi^2 j_{\ell'_1}( k_1 \chi )j_{\ell'_2}( k_2 \chi )j_{\ell'_3}( k_3 \chi )\,.
\eea
With the same manipulations as detailed in Section~3.4.1 of Ref.~\cite{DiDio:2015bua} we then obtain  for the $\HH^{-1} \partial_r^2v^{(2)}$ contribution to the bispectrum
\bea \label{bisp_RSD2}
b^{v^{(2)'}}_{\ell_1 \ell_2 \ell_3 } \left( z_1 , z_2 , z_3 \right)
&=& -\frac{16}{\pi^3} f(z_1) b_1(z_2) b_1(z_3)\!\! \int\!\! dk_1 dk_2 d k_3 k_1^2 k_2^2 k_3^2
G_2 \left( k_1, k_2, k_3 \right) P_R \left( k_2 \right) P_R \left( k_3
\right)
\nonumber \\
&&\times \
j''_{\ell_1} \left( k_1 r_1 \right) j_{\ell_2} \left( k_2 r_2 \right)
j_{\ell_3} \left( k_3 r_3 \right) T_\delta \left( k_2, z_1 \right)
T_\delta\left( k_3, z_1 \right) T_\delta \left( k_2, z_2 \right)
T_\delta \left( k_3 , z_3 \right)
\nonumber \\
&&
 \times\int_0^\infty d\chi \chi^2 j_{\ell_1} \left( k_1 \chi
\right)j_{\ell_2} \left( k_2 \chi \right)j_{\ell_3} \left( k_3 \chi
\right)  ~ + ~ \circlearrowleft\, ,
\eea
where $T_\delta \left( k, z \right) $ denotes the matter density
transfer function, $P_R \left( k \right)$ is the primordial curvature
power spectrum and $r_i=r(z_i)$ is the comoving distance to redshift
$z_i$. We assume transfer functions to be normalized to some initial
value of curvature perturbations, as in \cite{DiDio:2013bqa}. Given
the symmetry in the second and third indices of
eq.~(\ref{eq:rsd_corr}), there are only two more permutations (the
even ones) of the doublets $(\ell_i,z_i)$ to be added to the one shown
here and denoted by $\circlearrowleft$.

While in principle it is possible to reduce the integral by one
dimension by integrating analytically over $\chi$ as done in
Ref.~\cite{DiDio:2016gpd}, we show below that it is more convenient to
exchange the order of integration and express the result in terms of
Dirac deltas. However, as pointed out in
\cite{DiDio:2015bua,DiDio:2016gpd,Assassi:2017lea}, we stress that in
general care must be taken when exchanging the order of integration to
avoid UV divergences. While redshift binning (necessary in practice)
alleviates UV issues, divergences can be safely avoided only if
redshift bins are larger than the non-linearity scale, which does not
correspond to an optimal configuration for extracting the maximal
amount of cosmological information~\cite{DiDio:2013sea}, and it would
severely degrade the redshift accuracy of spectroscopic galaxy
surveys.

The main difference between the integrand of the bispectrum induced by RSD with respect to density perturbations (see Appendix~\ref{sec:b_dens_app}) comes from the second derivative of one spherical Bessel function.
Indeed, while the angular dependence in the kernel $G_2$, parametrized by the variable $k_1$, can be expressed by a few Legendre polynomials (as for the density perturbation in Sec.~\ref{sec:b_dens}), the second order derivative of the Bessel function introduces an additional factor $k_1^{-2}$. Because of this factor we cannot expand the part containing $G_2^{(0)}$ in a finite number of Legendre polynomials\footnote{See reference~\cite{Slepian:2018vds} for an alternative expansion.}.
To solve this problem we write
\be\label{e:rep-jpp}
j''_{\ell_1} \left( k_1 r_1 \right) = \int dx \ \delta_D (x- r_1 ) j''_{\ell_1} \left( k_1 x \right) = \int  dx \ k_1^{-2} \delta_D''(x-r_1) j_{\ell_1} \left( k_1 x \right) \, .
\ee
This also allows us to use the identity~\cite{Assassi:2017lea}
\be \label{identity_bessel}
\left(   \frac{\partial^2}{\partial \chi^2} + \frac{2}{\chi}
  \frac{\partial}{\partial \chi} - \frac{\ell \left( \ell +1 \right)
  }{\chi^2} \right) j_{\ell } \left( k \chi \right)  = -k^2 j_{\ell }
\left( k \chi \right) \;,
\ee
to evaluate the integral over the second derivative of the delta-Dirac
distribution. With this we can integrate analytically the products\footnote{The
Heaviside function is normalized such that $\Theta_H \left( 0 \right)
=1/2$. }
\bea
\label{eq20}
&&
\!\!\int \!\!dk_1  j_{\ell_1}  \left( k_1 x \right)   j_{\ell_1}\left( k_1 \chi \right) =
\frac{\pi}{2 \left( 1 +2 \ell_1\right)} \left( \chi^{-1-\ell_1} x^{\ell_1} \Theta_H \left( \chi- x \right) + x^{-1- \ell_1} \chi^{\ell_1} \Theta_H \left( x - \chi \right) \right)\, , \qquad
\\
&&\!\!\int \!\! dk_1 k_1^2 j_{\ell_1}  \left( k_1 x \right)   j_{\ell_1}\left( k_1 \chi \right) =
\frac{\pi}{2 x^2} \delta_D \left( \chi - x \right) \, ,
\\
&&
\!\!\int \!\! dk_1 k_1^4 j_{\ell_1} \left( k_1 x \right) j_{\ell_1} \left( k_1
  \chi \right) = \frac{\pi}{2 x^2} \left[ -
  \frac{\partial^2}{\partial \chi^2} - \frac{2}{\chi}
  \frac{\partial}{\partial \chi} + \frac{\ell_1 \left( \ell_1 +1 \right)
  }{\chi^2} \right] \delta_D \left( \chi - x \right)\, .
\eea
Using these identities as well as (\ref{e:rep-jpp}) we can perform the $k_1$ integrations of (\ref{bisp_RSD2}) and we find the following expression for the RSD contribution to the bispectrum
\bea \label{eq:b_rsd}
&& \hspace{-1cm}b^{v^{(2)'}}_{\ell_1 \ell_2 \ell_3 } \left( z_1 , z_2 , z_3 \right)
=
\nonumber \\
&&
 -2 \left( 4\pi \right)^2 f(z_1) b_1(z_2) b_1(z_3) \frac{1}{2 \ell_1 +1 } \!\!\int \!\! d\chi dx \frac{dk_2}{k_2} \frac{dk_3}{k_3} \chi^2  \delta''_D \left( x- r_1 \right)T_\delta \left( k_2, z_1\right) T_\delta \left( k_3, z_1\right)
\nonumber \\
&& \qquad  \qquad
T_\delta \left( k_2, z_2\right) T_\delta \left( k_3, z_3\right)
\mathcal{P}_R \left( k_2 \right) \mathcal{P}_R\left( k_3 \right)
j_{\ell_2 } \left( k_2 r_2 \right) j_{\ell_3 } \left( k_3 r_3 \right)
j_{\ell_2 } \left( k_2 \chi \right) j_{\ell_3 } \left( k_3 \chi
\right)
\nonumber \\
&&
\qquad \qquad
G^{(0)}_2 \left( k_2, k_3 \right) \left[ \chi^{-1-\ell_1} x^{\ell_1}
  \Theta_H \left( \chi- x \right) + x^{-1- \ell_1} \chi^{\ell_1}
  \Theta_H \left( x - \chi \right) \right]
\nonumber \\
&&
 -2 \left( 4\pi \right)^2 f(z_1) b_1(z_2) b_1(z_3) \!\!\int \!\!  dx \frac{dk_2}{k_2} \frac{dk_3}{k_3}  \delta''_D \left( x- r_1 \right)T_\delta \left( k_2, z_1\right) T_\delta \left( k_3, z_1\right) \nonumber \\
&& \qquad  \qquad\qquad\qquad
T_\delta \left( k_2, z_2\right) T_\delta \left( k_3, z_3\right)
\mathcal{P}_R \left( k_2 \right) \mathcal{P}_R\left( k_3 \right)
 \nonumber \\
&& \qquad  \qquad\qquad\qquad
j_{\ell_2 } \left( k_2 r_2 \right) j_{\ell_3 } \left( k_3 r_3 \right)
j_{\ell_2 } \left( k_2 x \right) j_{\ell_3 } \left( k_3 x \right) G^{(2)}_2 \left( k_2, k_3 \right)
\nonumber \\
&&
-2\left( 4\pi \right)^2 f(z_1) b_1(z_2) b_1(z_3) \!\!\int \!\! \frac{dx}{x^2} \frac{dk_2}{k_2} \frac{dk_3}{k_3}  \delta''_D \left( x - r_1 \right) T_\delta \left( k_2, z_1\right) T_\delta \left( k_3, z_1\right)  \nonumber \\
&& \qquad  \qquad\qquad\qquad
T_\delta \left( k_2, z_2\right) T_\delta \left( k_3, z_3\right)
\mathcal{P}_R \left( k_2 \right) \mathcal{P}_R\left( k_3 \right)
  \nonumber \\
&& \qquad  \qquad\qquad\qquad
D_{\ell_1}  \left[  j_{\ell_2 } \left( k_2 \chi \right) j_{\ell_3 } \left( k_3 \chi \right) { \chi^2} \right]_{\chi=x} j_{\ell_2 } \left( k_2 r_2 \right) j_{\ell_3 } \left( k_3 r_3 \right)
 G^{(4)}_2 \left( k_2, k_3 \right)
 \nonumber \\
 %%%%%%%%%%%%%%%%%
 &=& \hspace{0.2cm}
 2\left( 4\pi \right)^2 f(z_1) b_1(z_2) b_1(z_3) \!\!\int \!\!   \frac{dk_2}{k_2} \frac{dk_3}{k_3}   T_\delta \left( k_2, z_1\right) T_\delta \left( k_3, z_1\right) T_\delta \left( k_2, z_2\right) T_\delta \left( k_3, z_3\right)
\nonumber \\
&& \qquad  \qquad
\mathcal{P}_R \left( k_2 \right) \mathcal{P}_R\left( k_3 \right)
j_{\ell_2 } \left( k_2 r_2 \right) j_{\ell_3 } \left( k_3 r_3 \right)
j_{\ell_2 } \left( k_2 r_1 \right) j_{\ell_3 } \left( k_3 r_1 \right) G^{(0)}_2 \left( k_2, k_3 \right)
\nonumber \\
&&
%%%%%%
 -2\left( 4\pi \right)^2 f(z_1) b_1(z_2) b_1(z_3) \frac{1}{2 \ell_1 +1}  \int  d\chi \frac{dk_2}{k_2} \frac{dk_3}{k_3} \chi^2  T_\delta \left( k_2, z_1\right) T_\delta \left( k_3, z_1\right) T_\delta \left( k_2, z_2\right)
\nonumber \\
&& \qquad  \qquad
T_\delta \left( k_3, z_3\right)
\mathcal{P}_R \left( k_2 \right) \mathcal{P}_R\left( k_3 \right)
j_{\ell_2 } \left( k_2 r_2 \right) j_{\ell_3 } \left( k_3 r_3 \right)
j_{\ell_2 } \left( k_2 \chi \right) j_{\ell_3 } \left( k_3 \chi \right) G^{(0)}_2 \left( k_2, k_3 \right)
\nonumber \\
&&
\qquad  \qquad
\left[
\ell_1 \left( \ell_1 -1\right) \chi^{-1-\ell_1}  r_1^{-2+\ell_1} \Theta_H \left( \chi - r_1 \right)
+ \left( \ell_1 +1 \right) \left( \ell_1 +2 \right) r_1^{-3-\ell_1} \chi^{\ell_1} \Theta_H \left(  r_1 -\chi \right)
\right]
\nonumber \\
&&
%%%%%%
 -2\left( 4\pi \right)^2 f(z_1) b_1(z_2) b_1(z_3) \!\!\int \!\!   \frac{dk_2}{k_2} \frac{dk_3}{k_3}  T_\delta \left( k_2, z_1\right) T_\delta \left( k_3, z_1\right) T_\delta \left( k_2, z_2\right) T_\delta \left( k_3, z_3\right)
\nonumber \\
&& \qquad  \qquad
\mathcal{P}_R \left( k_2 \right) \mathcal{P}_R\left( k_3 \right)
j_{\ell_2 } \left( k_2 r_2 \right) j_{\ell_3 } \left( k_3 r_3 \right)
\frac{\partial^2}{\partial r_1^2} \left( j_{\ell_2 } \left( k_2 r_1 \right) j_{\ell_3 } \left( k_3 r_1 \right)\right) G^{(2)}_2 \left( k_2, k_3 \right)
\nonumber \\
&&
%%%%%%
-2\left( 4\pi \right)^2 f(z_1) b_1(z_2) b_1(z_3) \!\!\int \!\!  \frac{dk_2}{k_2} \frac{dk_3}{k_3} \frac{1}{r_1^2}  \mathcal{P}_R \left( k_2 \right) \mathcal{P}_R\left( k_3 \right)
\nonumber \\
&& \qquad  \qquad\qquad\qquad
 T_\delta \left( k_2, z_1\right) T_\delta \left( k_3, z_1\right)  T_\delta \left( k_2, z_2\right) T_\delta \left( k_3, z_3\right) G^{(4)}_2 \left( k_2, k_3 \right)
\nonumber \\
&& \qquad  \qquad\qquad\qquad
\frac{\partial^2}{\partial r_1^2} \left(D_{\ell_1}  \left[  j_{\ell_2 } \left( k_2 \chi \right) j_{\ell_3 } \left( k_3 \chi \right) { \chi^2} \right]_{\chi=r_1} \right) j_{\ell_2 } \left( k_2 r_2 \right) j_{\ell_3 } \left( k_3 r_3 \right)
  \, ,
 \eea
 where we have introduced the dimensionless power spectrum
$\mathcal{P}_R(k) = \frac{k^3}{2\pi^2} P_R(k)$ and defined the
differential operator
\be
D_\ell = - \frac{\partial^2}{\partial \chi^2} + \frac{2}{\chi} \frac{\partial}{\partial \chi} + \frac{\ell \left( \ell + 1\right) -2}{\chi^2} \, .
\ee
For the second equal sign we have also performed the $x$-integration and an integration by parts on the $G^{(0)}_2$ term. For the second and the fourth terms we could perform both the $x$ and the $\chi$ integrations. The first term still contains a triple integral.  The terms proportional to $G^{(2)}_2$ and $G^{(4)}_2$ are simply sums of products of 1-dimensional integrals.

Summing up the previous expression we obtain
\bea
\label{v2_eq}
&b^{v^{(2)'}}_{\ell_1 \ell_2 \ell_3 }& \left( z_1 , z_2 , z_3 \right)
= 2\left( 4\pi \right)^2 f(z_1) b_1(z_2) b_1(z_3) \int
\frac{dk_2}{k_2} \frac{dk_3}{k_3} \mathcal{P}_R \left( k_2 \right)
\mathcal{P}_R\left( k_3 \right)
\nonumber \\
&&T_\delta \left( k_2, z_1\right) T_\delta \left( k_3,
  z_1\right) T_\delta \left( k_2, z_2\right) T_\delta \left( k_3,
  z_3\right) j_{\ell_2 } \left( k_2 r_2 \right) j_{\ell_3 } \left( k_3
  r_3 \right)
\nonumber \\
&&\Bigg[  j_{\ell_2 } \left( k_2 r_1 \right) j_{\ell_3 } \left(
  k_3 r_1 \right) G^{(0)}_2 \left( k_2, k_3 \right)
\nonumber \\
&& - \frac{1}{2\ell_1 + 1} \int d\chi \left( \ell_1
  \left( \ell_1 -1\right) \frac{r_1^{\ell_1-2}}{\chi^{\ell_1-1}}
  \Theta_H \left( \chi - r_1 \right) \right.
\nonumber \\
&&\qquad \left. + \left( \ell_1 +1 \right) \left(
    \ell_1 +2 \right) \frac{\chi^{\ell_1+2}}{r_1^{3+\ell_1}} \Theta_H
  \left( r_1 -\chi \right) \right) j_{\ell_2 } \left( k_2
  \chi \right) j_{\ell_3 }\left( k_3 \chi \right) G^{(0)}_2\left( k_2, k_3 \right)
\nonumber \\
&& -\frac{\partial^2}{\partial r_1^2} \left( j_{\ell_2 } \left( k_2
    r_1 \right) j_{\ell_3 } \left( k_3 r_1 \right)\right) G^{(2)}_2
\left( k_2, k_3 \right)
\nonumber \\
&& - \frac{1}{r_1^2} \frac{\partial^2}{\partial r_1^2} \left(D_{\ell_1}
  \left[ j_{\ell_2 }( k_2 r_1) j_{\ell_3 }( k_3
      r_1) {r_1^2} \right]\right)G^{(4)}_2( k_2, k_3) \Bigg]\, .
\eea

In order to add also the RSD in the linear term, we simply replace
\be \label{include_RSD}
b_1(z_i) T_\delta \left(k_i , z_i \right) j_{\ell_i} \left( k_i r_i \right)
\rightarrow b_1(z_i) T_\delta\left(k_i , z_i \right) j_{\ell_i} \left(
  k_i r_i \right) -f(z_i) T_\delta \left( k_i , z_i
\right) j''_\ell \left( k_i r_i \right) \, .
\ee
The important point for numerical efficiency when evaluating
$b^{v^{(2)'}}_{\ell_1 \ell_2 \ell_3 }$ is that the kernels
$G_2^{(i)} \left( k_2 , k_3 \right) $, with $i=0,2,4$, are separable
in $k_2$ and $k_3$. Because of that we can reduce the dimensionality
of the integrals of eq.~\eqref{v2_eq}. In Appendix~\ref{app:rsd} we
rewrite eq.~\eqref{v2_eq} explicitly as a sum of products of
1-dimensional integrals, in terms of generalized spectra defined in eq.~\eqref{eq:c_ll}, except for the terms involving an additional
  integral along $\chi$. Indeed, naively exchanging the integration
  order for these latter terms would lead to divergences for a
  $\Lambda $CDM cosmology due to the powers of $k_2$ and $k_3$ in the
  kernel $G_2^{(0)}$.

%%%%%%%%%%%%%
\subsection{Density $\delta^{(2)}$}
\label{sec:b_dens}
The second term we consider is the second order density perturbation.
In order to have a consistent bias expansion we consider also the local bias parameters $b_1$, $b_2$ and $b_s$
\bea \label{eq:dens_corr}
&& B^{\delta^{(2)}}(\bn_1,\bn_2,\bn_3,z_1,z_2,z_3) ~~ = \nonumber\\
&&
\Bigg\langle \left( b_1(z_1) \delta^{(2)}\left( \bn_1, z_1 \right) +
\frac{b_2(z_1)}{2} (\delta^{(1)}\left( \bn_1, z_1 \right))^2 +
b_s(z_1) s^2\left( \bn_1, z_1 \right) \right)
\nonumber \\
&&
\qquad
\times b_1(z_2) \delta^{(1)}
\left(
\bn_2, z_2 \right) b_1(z_3) \delta^{(1)} \left( \bn_3, z_3 \right)
+ \circlearrowleft \Bigg\rangle
\eea
with
\be \label{eq:dens2_expr}
\delta^{(2)} \left( \bk ,z \right) = \frac{1}{\left( 2 \pi \right)^3} \int d^3k_1 d^3k_2 \delta_D \left( \bk - \bk_1 - \bk_2 \right) F_2 \left( \bk_1 , \bk_2 \right) \delta \left( \bk_1 , z \right)
\delta \left( \bk_2 , z \right)
\ee
and
\bea
\label{def_F2}
 F_2 \left( \bk_2 , \bk_3 \right) &=& \frac{5}{7} + \frac{1}{2} \frac{\bk_2 \cdot \bk_3 }{k_2 k_3} \left( \frac{k_2}{k_3} + \frac{k_3}{k_2} \right) + \frac{2}{7} \left( \frac{\bk_2 \cdot \bk_3}{k_2 k_3} \right)^2
 \nonumber \\
 &=& \frac{17}{21} + \frac{1}{2} \left( \frac{k_2}{k_3} + \frac{k_3}{k_2} \right)P_1 \left( \hat\bk_2 \cdot \hat\bk_3 \right) + \frac{4}{21} P_2  \left( \hat\bk_2 \cdot \hat\bk_3 \right) \, .
\eea

Again, we follow \cite{DiDio:2015bua}, where the bispectrum is computed
separately for the monopole $b^{\delta0}_{\ell_1 \ell_2 \ell_3}$,
dipole $b^{\delta1}_{\ell_1 \ell_2 \ell_3}$ and quadrupole
$b^{\delta2}_{\ell_1 \ell_2 \ell_3}$ terms for the case $b_1=1$,
$b_2=b_s=0$, where the multipole expansion refers to the Legendre polynomials in eq.~\eqref{def_F2}.
Indeed, it is trivial to generalize this result. We denote the term
proportional to $b_2/2$ by $ b^{\delta^2}_{\ell_1 \ell_2 \ell_3}$ and
the tidal term proportional to $b_s$ by $b^{s^2}_{\ell_1 \ell_2
  \ell_3}$. The term proportional to $b_2/2$ is just a monopole with pre-factor one, hence
\be
\label{bias_b2_monopole} b^{\delta^2}_{\ell_1 \ell_2 \ell_3}
\left( z_1 , z_2 , z_3 \right) = \frac{21}{17} b^{\delta0}_{\ell_1
  \ell_2 \ell_3} \left( z_1 , z_2 , z_3 \right)
\ee
while the tidal
term becomes\footnote{ In Fourier space the tidal term can be expressed as
$$
s^2 \left( \bk , z \right) = \frac{1}{(2\pi)^3}\int d^3k_1 d^3k_2
  \delta_D \left( \bk - \bk_1 - \bk_2 \right) S_2 \left( \bk_1 , \bk_2
  \right) \delta \left( \bk_1 , z \right) \delta \left( \bk_2 , z \right)
$$
where
$$
S_2 \left( \bk_1 , \bk_2 \right) = -\frac{1}{3} + \frac{\left( \bk_1 \cdot \bk_2 \right)^2}{ k_1^2 k_2^2} = \frac{2}{3} P_2 \left( \hat \bk_1 \cdot \hat \bk_2 \right)
$$
Comparing with the quadrupole of eq.~\eqref{def_F2} we obtain eq.~\eqref{bias_b2_quadrupole}.
}
\be\label{bias_b2_quadrupole}
b^{s^2}_{\ell_1 \ell_2 \ell_3} \left( z_1 , z_2 , z_3 \right) = \frac{7}{2} b^{\delta2}_{\ell_1 \ell_2 \ell_3} \left( z_1 , z_2 , z_3 \right)
\ee

We define generalized angular power spectra as in~\cite{DiDio:2015bua}:
\be\label{eq:c_ll}
\prescript{n}{}{c}^{AB}_{\ell\ \ell'}(z_1,z_2)= i^{\ell-\ell'} 4\pi
\int \frac{dk}{k} k^n \mathcal{P}_R \left( k \right) \Delta^A_\ell \left( k , r_1 \right) \Delta^B_{\ell'} \left( k , r_2 \right)\,.
\ee
Here $\Delta^A_\ell \left( k , r\right)$ is the angular transfer function related to the perturbation $A$. In particular we need (using the same notation as~\cite{DiDio:2013bqa,DiDio:2015bua,DiDio:2016ykq,DiDio:2016gpd})
\bea \label{transferT}
\Delta^{\De}_\ell \left( k , r \right) &=& b_1(r) T_\delta \left( k , r \right) j_\ell \left( k r \right) + \frac{k}{\HH}T_V \left( k , r \right) j''_\ell \left( k r \right) \, , \\
\label{transfer_delta}
\Delta^{\delta}_\ell \left( k , r \right) &=& T_\delta \left( k , r \right) j_\ell \left( k r \right)
 \, , \\
 \label{transfer_drdelta}
\Delta^{\delta'}_\ell \left( k , r \right) &=&\frac{k}{\HH} T_\delta \left( k , r \right) j'_\ell \left( k r \right)  \, ,
\\
\Delta^{v}_\ell \left( k , r \right) &=& T_V \left( k , r \right) j'_\ell \left( k r \right) \, , \\
\Delta^{v'}_\ell \left( k , r \right) &=& \frac{k}{\HH} T_V \left( k , r \right) j''_\ell \left( k r \right) \, , \\
\Delta^{v''}_\ell \left( k , r \right) &=& \left( \frac{k}{\HH}\right)^2T_V \left( k , r \right) j'''_\ell \left( k r \right) \, ,
\eea
where $T_\delta(k,z)$ denotes the linear transfer function for the density $\delta$ and for the velocity we have
$T_V(k,z)=-(\HH(z)/k) f(z)T_\delta(k,z)$.
For $n=0$ we shall drop this pre-superscript in $\prescript{n}{}{c}^{AB}_{\ell\ \ell'}$ and if $\ell'=\ell$ we just indicate it by one subscript $\ell$ so that e.g. $\prescript{0}{}{c}^{AB}_{\ell\ \ell}\equiv {c}^{AB}_{\ell}$.

With this notation we can write
\begin{eqnarray}
 b^{\delta^{(2)}}_{\ell_1,\ell_2,\ell_3}(z_1,z_2,z_3)
&=&\left( b_1\left( z_1 \right) + \frac{21}{34} b_2 \left( z_1 \right) \right) b^{\De  \ \delta0 }_{\ell_1 \ell_2 \ell_3 } \left( z_1 , z_2 , z_3 \right)
 + b_1 \left( z_1 \right)  b^{\De  \ \delta1}_{\ell_1 \ell_2 \ell_3 }\left( z_1 , z_2 , z_3 \right)
 \nonumber \\
&& +
 \left( b_1\left( z_1 \right) + \frac{7}{2} b_s \left( z_1 \right)
 \right) b^{\De  \ \delta2}_{\ell_1 \ell_2 \ell_3 } \left( z_1 , z_2 ,
   z_3 \right)  {+\circlearrowleft}\;,
\end{eqnarray}
where we introduce
\begin{itemize}
\item Monopole:
\be
 b^{ \delta0 }_{\ell_1 \ell_2 \ell_3 } \left( z_1 , z_2 , z_3 \right)
 =
 \frac{34}{21}
c_{\ell_1}^{\delta\De}(z_1,z_2) c_{\ell_2}^{\delta\De}(z_1,z_3)
\,.
\ee

\item Dipole:
\bea \label{eq:b_dens_dip}
\hspace*{-1cm} b^{ \delta1 }_{\ell_1 \ell_2 \ell_3 } \left( z_1 , z_2 , z_3 \right)
 &=&
 \frac{ \left(g_{\ell_1\ell_2\ell_3}\right)^{-1}}{16\pi^2} \sum_{\ell'\ell''} (2\ell'+1)(2\ell''+1) Q_{1\ \ell' \ell''}^{\ell_1 \ell_2 \ell_3}
\nonumber \\
&& \hspace{-3cm} \times \left[
  \prescript{1}{}{c}^{\delta \De}_{\ell'' \ell_2}(z_1,z_2)
  \prescript{-1}{}{c}^{\delta \De}_{\ell' \ell_3}(z_1,z_3)
  +
  \prescript{-1}{}{c}^{\delta \De}_{\ell'' \ell_2}(z_1,z_2)
  \prescript{1}{}{c}^{\delta \De}_{\ell' \ell_3}(z_1,z_3)
  \right]
\eea
The geometrical factors $g_{\ell_1\ell_2\ell_3}$ and
$Q_{\ell\ \ell'\ell''}^{\ell_1\ell_2\ell_3}$ are defined in Appendix
\ref{sec:gener-spectra-geom}. The quantity
$Q_{1\ \ell'\ell''}^{\ell_1\ell_2\ell_3}$ is zero unless
$\ell'=\ell_2 \pm 1$ and $\ell''=\ell_1 \pm 1$ so that
$i^{\ell'+\ell''} (-i)^{\ell_1+\ell_2} = \pm 1$.
\
\item Quadrupole:
\be
  b^{\delta2 }_{\ell_1 \ell_2 \ell_3 } \left( z_1 , z_2 , z_3 \right)
 =
\frac{\left(g_{\ell_1\ell_2\ell_3}\right)^{-1}}{42\pi^2}
\sum_{\ell'\ell''} (2\ell'+1)(2\ell''+1) Q_{2\ \ell' \ell''}^{\ell_1 \ell_2 \ell_3} \
{c}^{\delta \De}_{\ell'' \ell_2}(z_1,z_2) \  {c}^{\delta \De}_{\ell' \ell_3}(z_1,z_3)
 \label{eq:b_dens_quad}
\ee
where $Q_{2\ \ell'\ell''}^{\ell_1\ell_2\ell_3}$ is zero unless
$\ell'=\ell_2\pm 2, \ell_2$ and $\ell''=\ell_1\pm 2, \ell_1$. We then
have again $i^{\ell'+\ell''} (-i)^{\ell_1+\ell_2} = \pm 1$.
\end{itemize}

In Appendix~\ref{sec:b_dens_app} we also present an alternative derivation
of the density bispectrum  using the approach adopted in the
previous section for RSD.

%%%%%%%%
\subsection{Products of linear terms}
Using Wick's theorem, one finds that all the terms in the second line of eq.~\eqref{2nd_counts} induce a bispectrum which is the sum of products of certain power spectra. See~\cite{DiDio:2015bua,DiDio:2016gpd} for an explicit computation.
In detail we find
\bea
%%%%%%%%%%%%%%%%%%%
b^{\delta v'}_{\ell_1 \ell_2 \ell_3} \left( z_1 , z_2 , z_3 \right)&=&
c^{\delta_g \De}_{\ell_2} \left( z_1 , z_2 \right) c^{v' \De}_{\ell_3} \left( z_1 , z_3 \right)
+
c^{v' \De}_{\ell_2} \left( z_1 , z_2 \right) c^{\delta_g \De}_{\ell_3} \left( z_1 , z_3 \right)  + \circlearrowleft\, , \quad
\\
%%%%%%%%%%%%%%
b^{v'^2}_{\ell_1 \ell_2 \ell_3} \left( z_1 , z_2 , z_3 \right)
&=&
 2 c^{v' \De}_{\ell_2} \left( z_1 , z_2 \right) c^{v' \De}_{\ell_3} \left( z_1 , z_3 \right)  + \circlearrowleft \, ,
\\
%%%%%%%%%%%%%%
b^{\delta' v}_{\ell_1 \ell_2 \ell_3} \left( z_1 , z_2 , z_3 \right)&=&
b_1 \left( z_1 \right) c^{\delta' \De}_{\ell_2} \left( z_1 , z_2 \right) c^{v \De}_{\ell_3} \left( z_1 , z_3 \right)
\nonumber \\
&&
+
b_1 \left( z_1 \right)
c^{v \De}_{\ell_2} \left( z_1 , z_2 \right) c^{\delta'
  \De}_{\ell_3} \left( z_1 , z_3 \right)  + \circlearrowleft\, ,
\\
%%%%%%%%%%%%%%
b^{v'' v}_{\ell_1 \ell_2 \ell_3} \left( z_1 , z_2 , z_3 \right)&=&
c^{v'' \De}_{\ell_2} \left( z_1 , z_2 \right) c^{v \De}_{\ell_3} \left( z_1 , z_3 \right)
+
c^{v \De}_{\ell_2} \left( z_1 , z_2 \right) c^{v'' \De}_{\ell_3} \left( z_1 , z_3 \right)  + \circlearrowleft \, .
\eea
 The two additional permutations are the even ones of the doublets
 $(\ell_i,z_i)$.

\subsection{The full Bispectrum}

Adding all the terms computed in the previous sections, we obtain the leading galaxy bispectrum  as
\bea \label{eq:full_bisp}
b_{\ell_1 \ell_2 \ell_3 } \left( z_1 , z_2 , z_3 \right) &=& \ b^{\delta^{(2)}}_{\ell_1,\ell_2,\ell_3}(z_1,z_2,z_3)+ b^{ v^{(2)'}}_{\ell_1 \ell_2 \ell_3 } \left( z_1 , z_2 , z_3 \right)
  \nonumber \\
&+&
b^{\delta v'}_{\ell_1 \ell_2 \ell_3} \left( z_1 , z_2 , z_3 \right)
+ b^{ v'^2}_{\ell_1 \ell_2 \ell_3} \left( z_1 , z_2 , z_3 \right)
  \nonumber \\
&+&
b^{\delta' v}_{\ell_1 \ell_2 \ell_3} \left( z_1 , z_2 , z_3 \right)
+ b^{v'' v}_{\ell_1 \ell_2 \ell_3} \left( z_1 , z_2 , z_3 \right) \, .
 \eea

%%%%%%%%%%%%%%%%%%%%%%%%%%%%%%%%%%%%%%%%%%%%%%%%%
\section{Numerical results}
\label{sec:numerical}

\subsection{The different contributions to the bispectrum}
In this section we show the numerical redshift dependent angular
bispectrum for different configurations. All the results are obtained
 with the following cosmological parameters: $h=0.67$,
$\Omega_b= 0.05 $, $\Omega_{cdm} =0.27$, and consistently with
  the previous section vanishing curvature. The
amplitude of the primordial curvature power spectrum is set to
$A_s=2.3 \times 10^{-9}$, the pivot scale is
$k_{\rm pivot}=0.05 {\rm Mpc}^{-1}$, the spectral index is $n_s=0.962$
and we assume no running. As discussed in Section
  \ref{sec:angbisp}, bispectra are computed correlating a second-order
  term with two first-order terms, where the latter include density
  and RSD. Clustering bias (figure \ref{f:bias}) is always included.

  Our numerical computations rely on {\sc byspectrum},\footnote{The
    code is found at \url{https://gitlab.com/montanari/byspectrum}.} a
  Python code making use of a low-level C++ library for the
  computation of generalized spectra that wraps the \class{}
  code~\cite{Blas:2011rf,DiDio:2013bqa} to retrieve the transfer
  functions.

\begin{figure}[t]
\begin{center}
\includegraphics[width=0.45\textwidth]{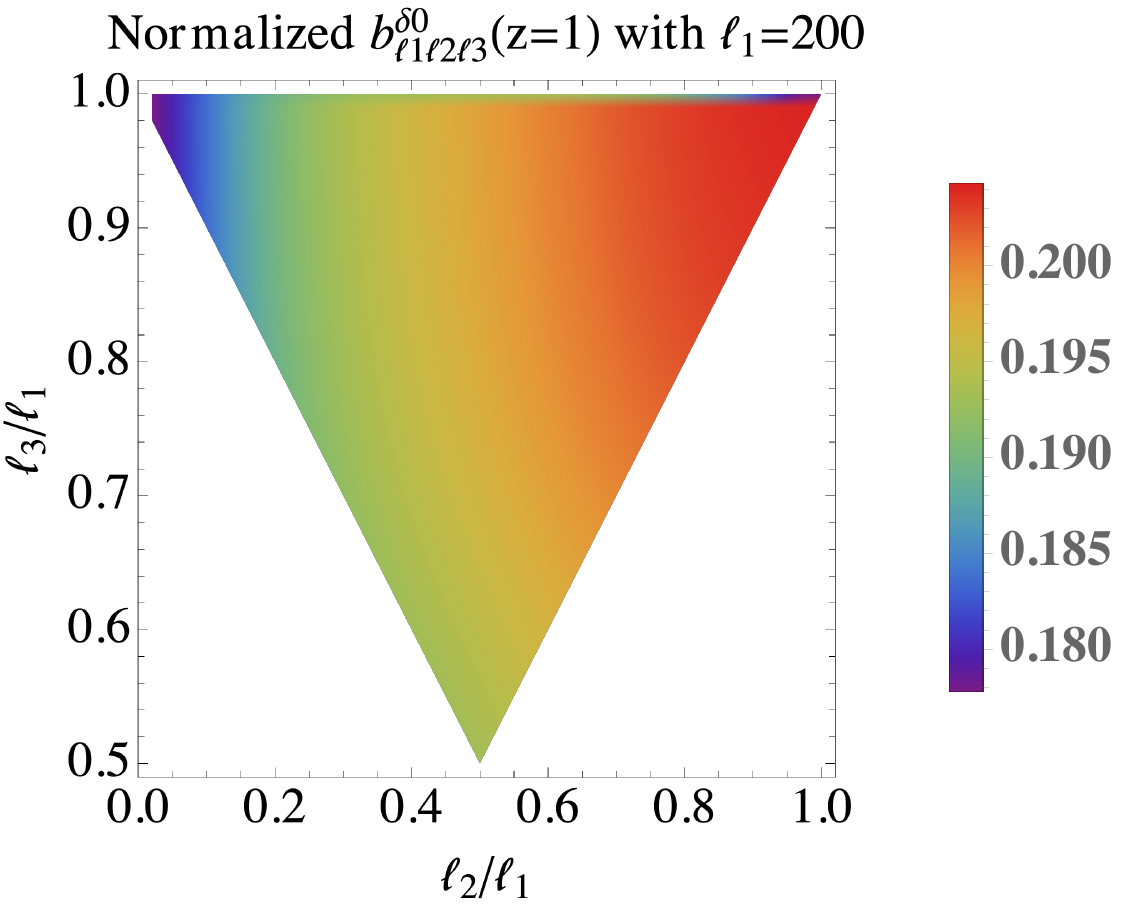}
\includegraphics[width=0.45\textwidth]{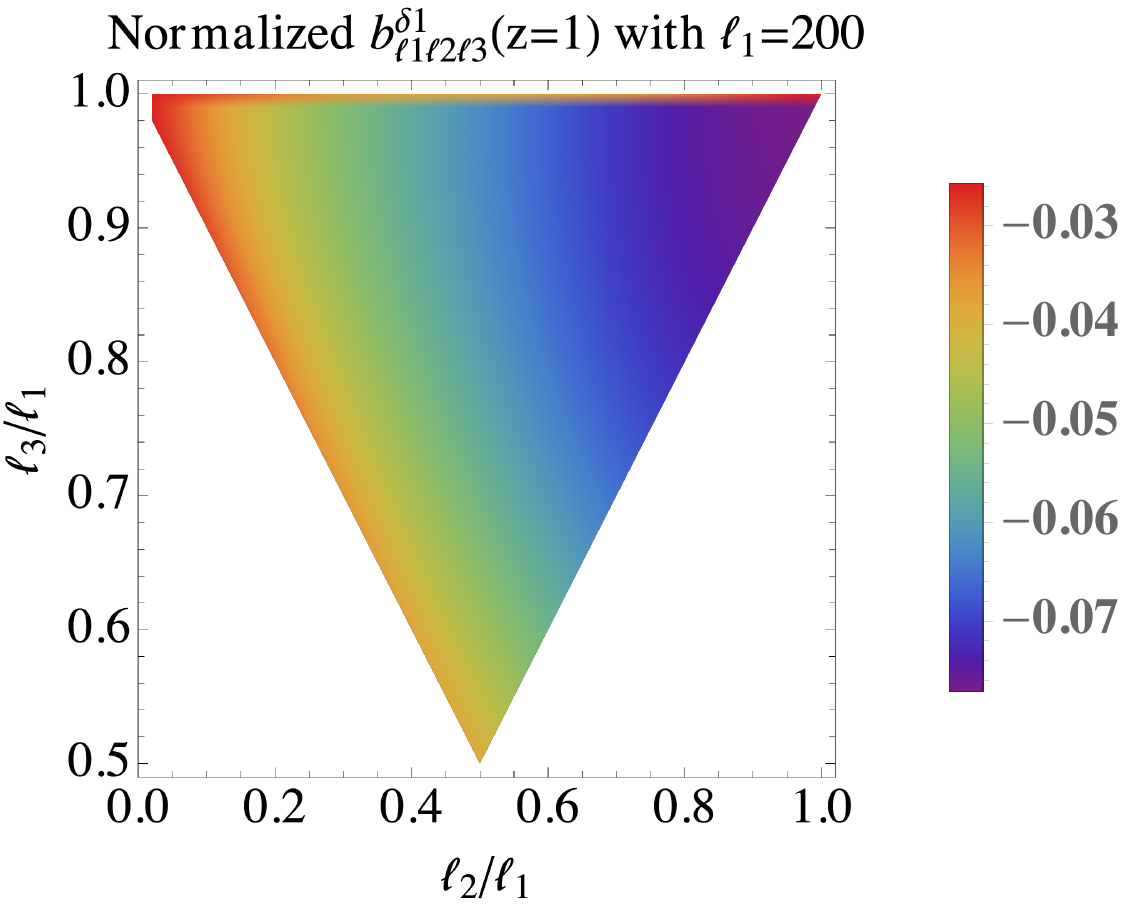}
\\
\includegraphics[width=0.45\textwidth]{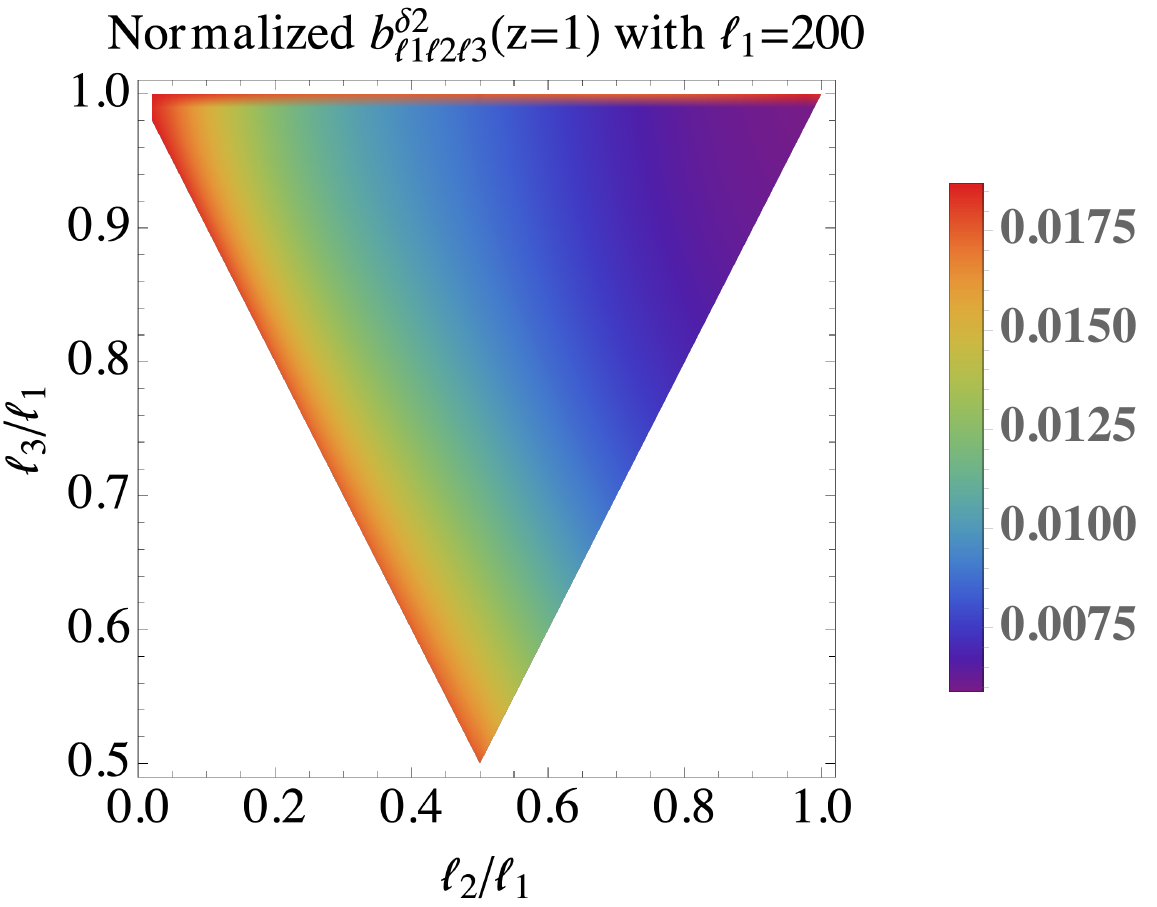}
\includegraphics[width=0.45\textwidth]{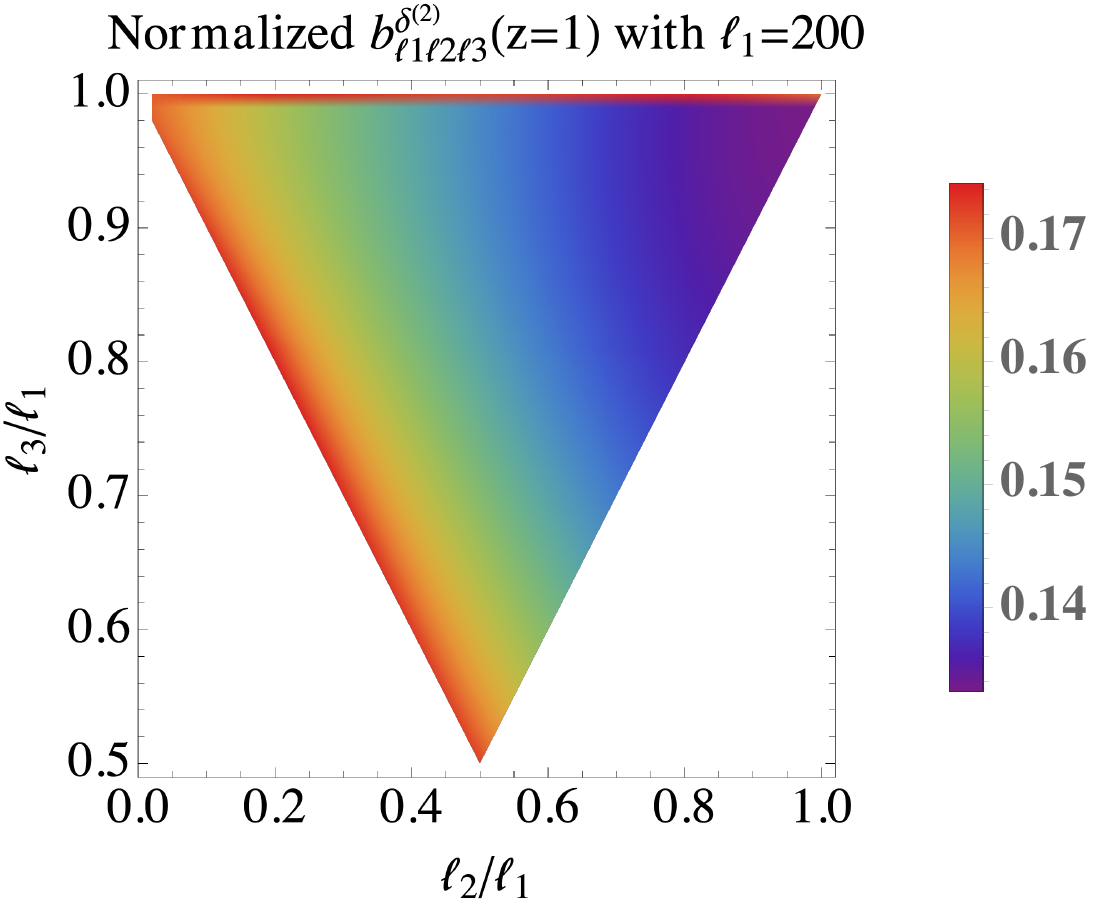}
\caption{Reduced bispectrum of the density term $\delta^{(2)}$
  normalized to the square of power spectra. We set
  $z_1=z_2=z_3=1$. Note that $\ell_1=200$ at $z=1$ corresponds to
    a comoving scale of about 100 Mpc.}
\label{Fig_den}
\end{center}
\end{figure}
%%%%%%%%%%%%%%%%%%%%%%%%
\begin{figure}[t]
\begin{center}
\includegraphics[width=0.45\textwidth]{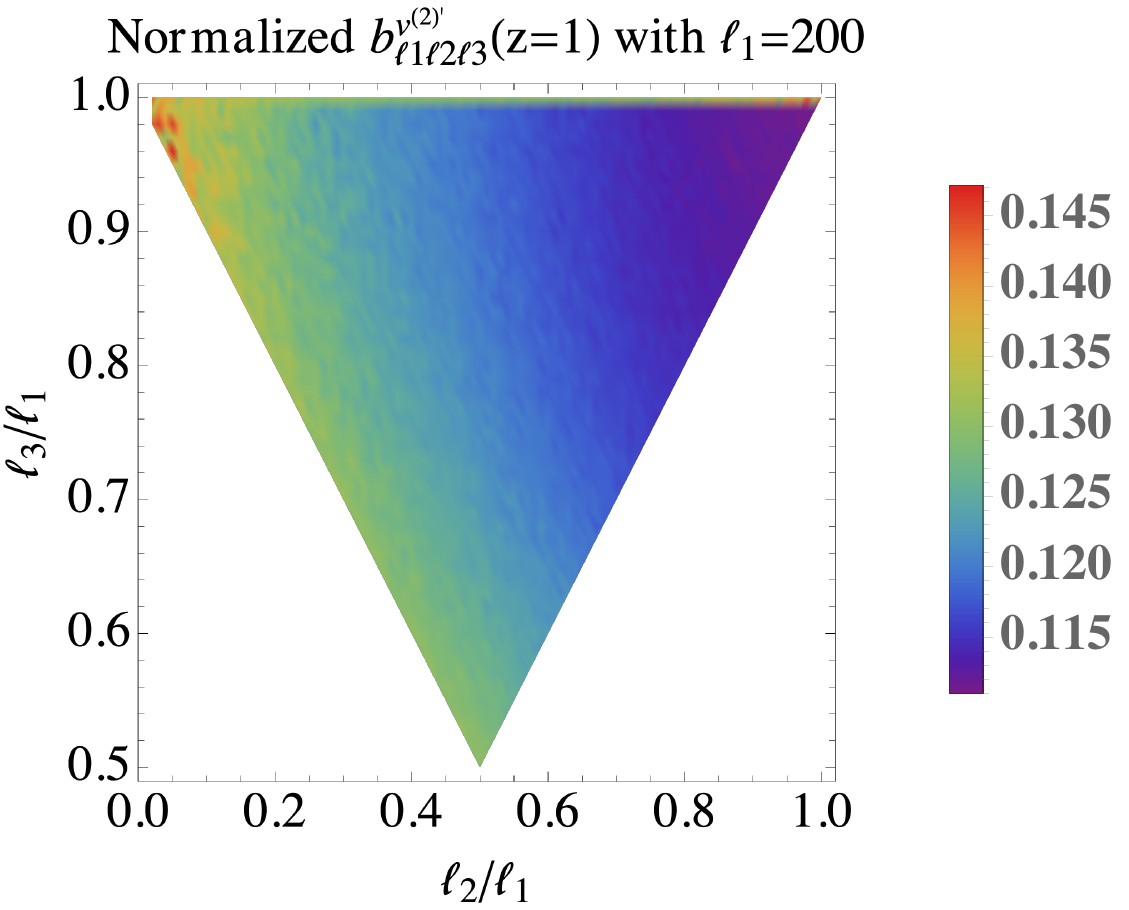}
\caption{Normalized reduced bispectrum of the RSD term
  $\HH^{-1} \partial^2_r v^{(2)}$. We set $z_1=z_2=z_3=1$.}
\label{fig:triangle_rsd}
\end{center}
\end{figure}
%%%%%%%%%%%%%%%%%%%%%%%%
\begin{figure}[t]
\begin{center}
\includegraphics[width=0.45\textwidth]{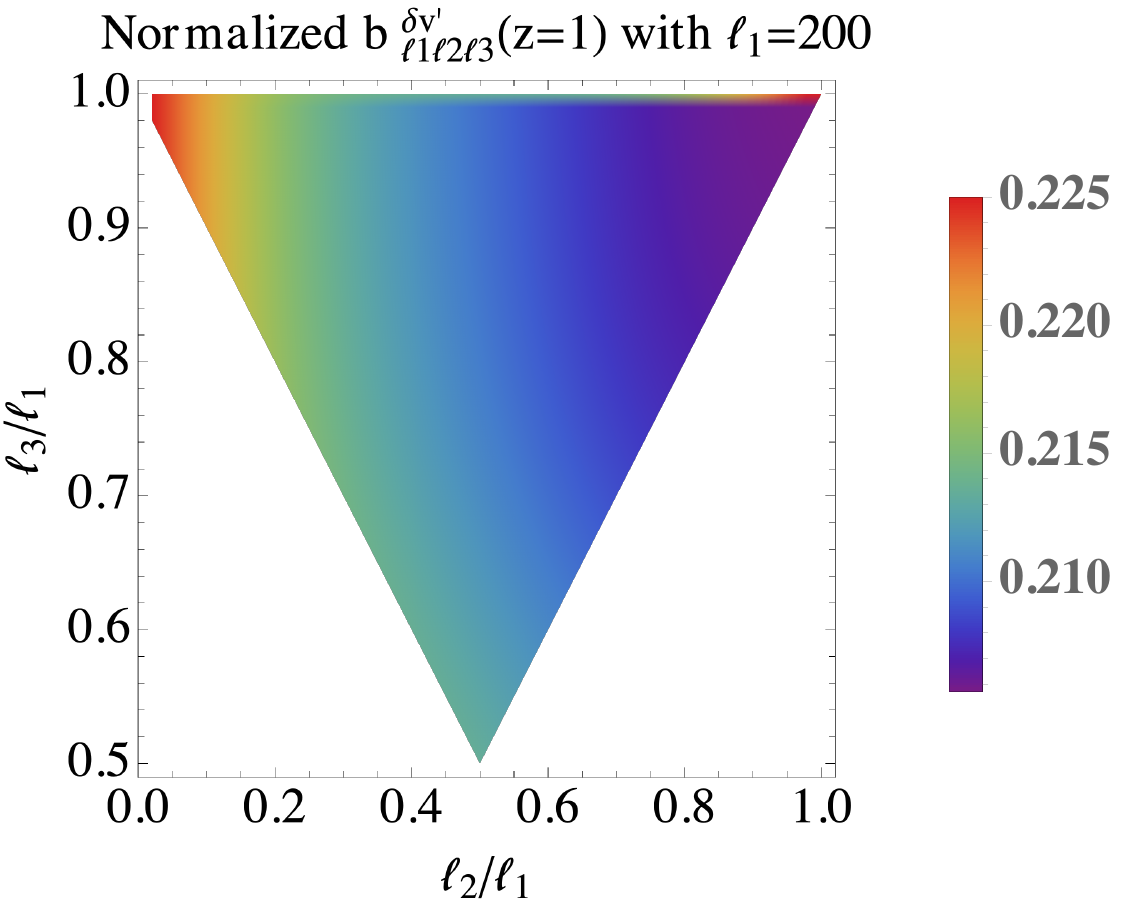}
\includegraphics[width=0.45\textwidth]{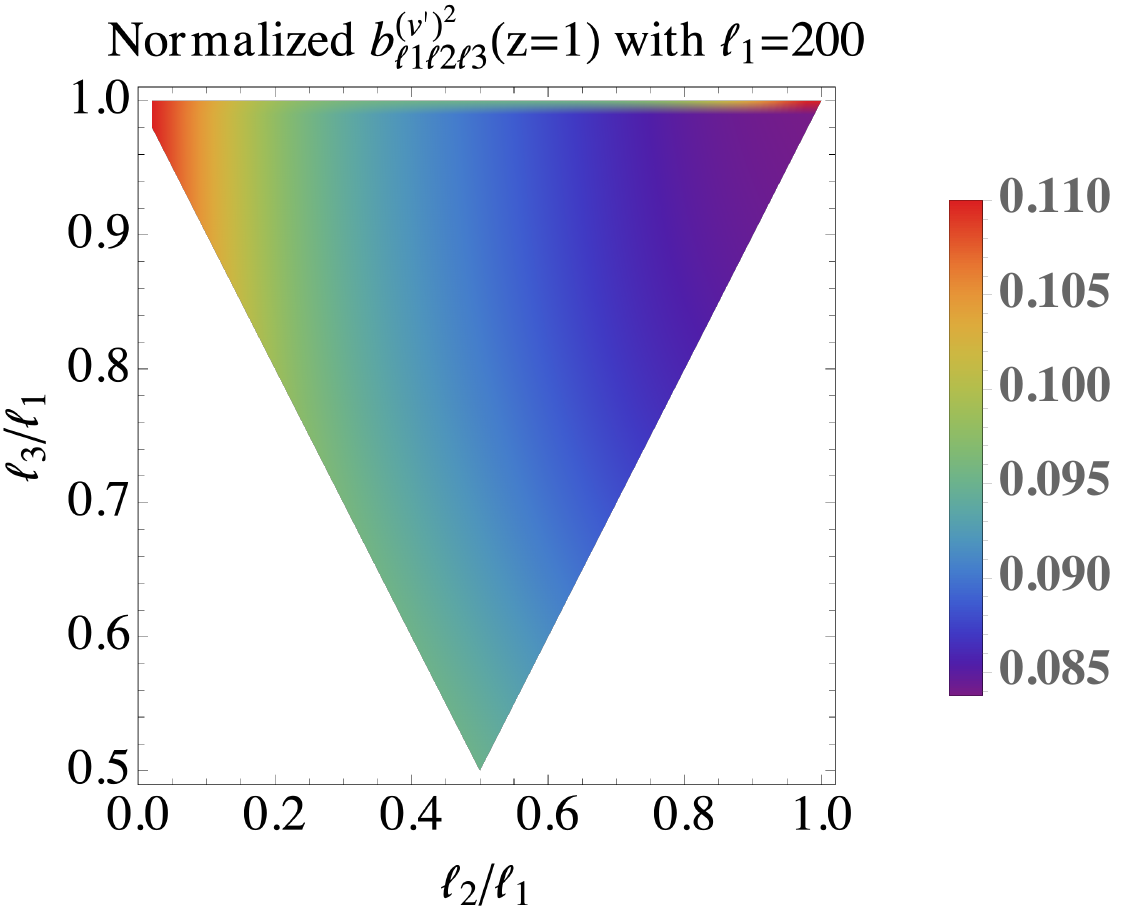}
\\
\includegraphics[width=0.45\textwidth]{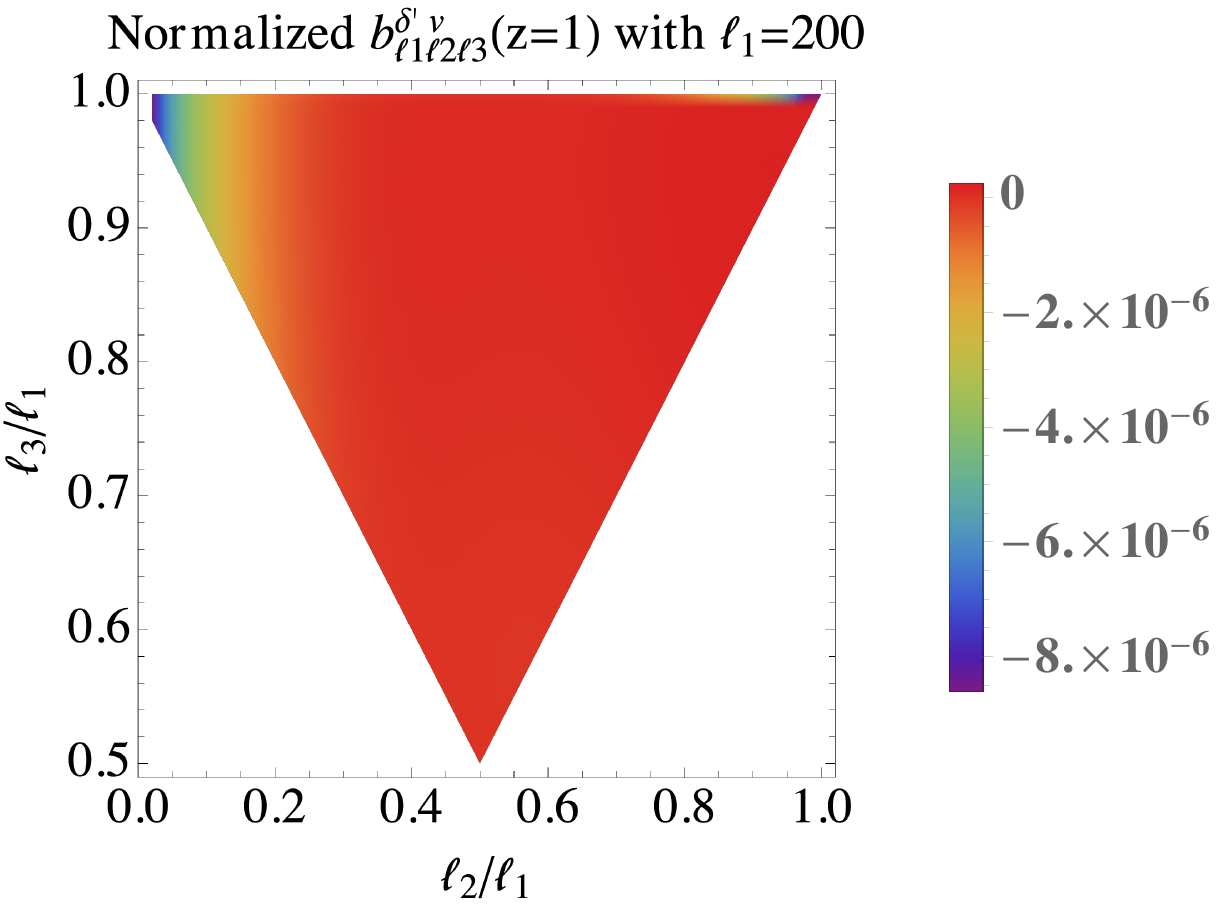}
\includegraphics[width=0.45\textwidth]{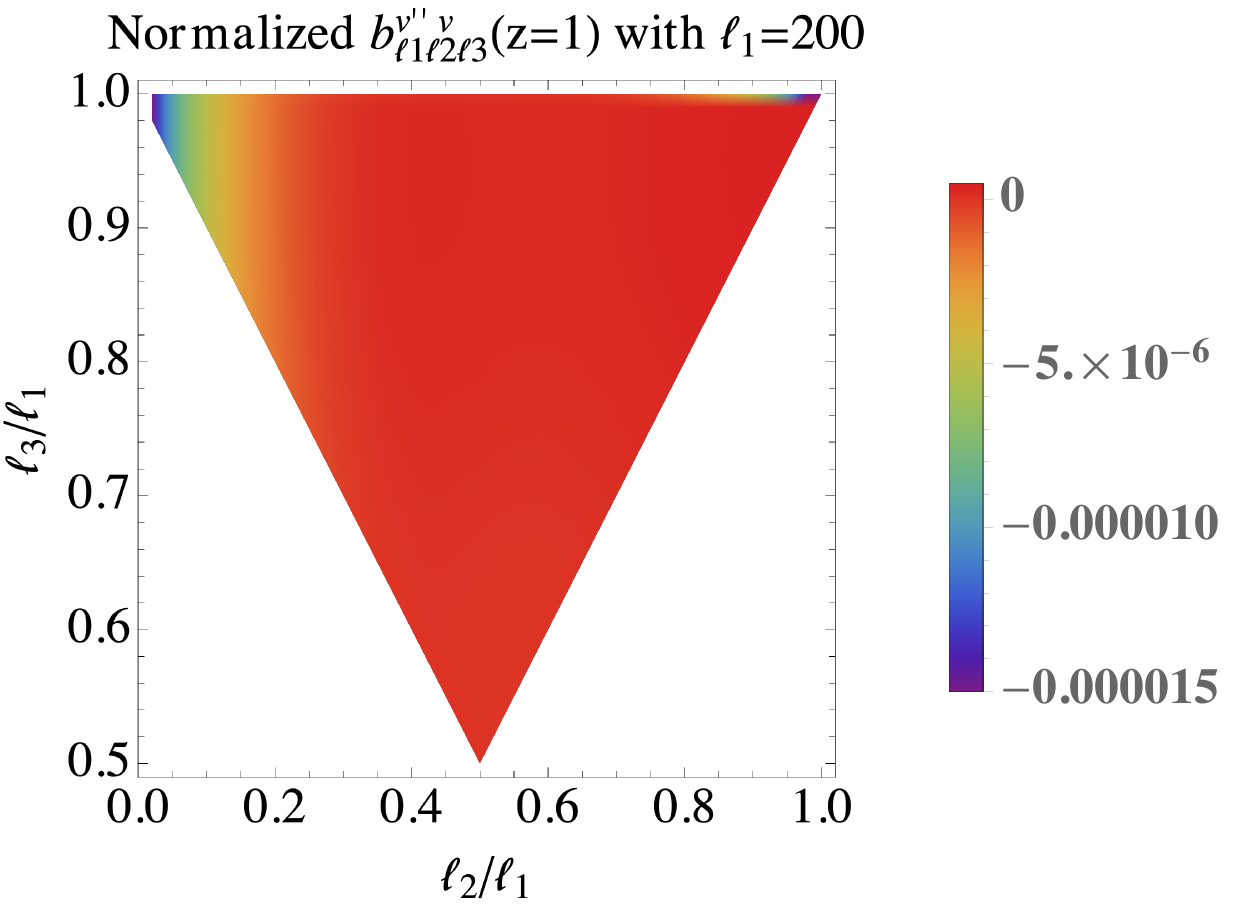}
\caption{Normalized reduced bispectrum of terms involving a
    velocity contribution (other than RSD). The amplitude of
    the bottom panels is strongly suppressed due to  spherical
    Bessel functions which are out of phase. We set
    $z_1=z_2=z_3=1$.}
\label{Fig_den_other}
\end{center}
\end{figure}

In figures~\ref{Fig_den}-\ref{Fig_den_other}, we plot the ratio between
the bispectrum contributions normalized with respect to the
square of the power spectra (which include density and RSD perturbations), i.e.
\be
\label{renormalized_bisp}
\frac{b^{\rm{term}}_{\ell_1 \ell_2 \ell_3 } \left( z_1 ,z_2 , z_3 \right)}{c_{\ell_2}^{\Delta \Delta} \left( z_1, z_2 \right) c_{\ell_3}^{\Delta \Delta} \left( z_1, z_3 \right)   +c_{\ell_1}^{\Delta \Delta} \left( z_2, z_1 \right) c_{\ell_3}^{\Delta \Delta} \left( z_2, z_3 \right) +c_{\ell_1}^{\Delta \Delta}\left( z_3, z_1 \right) c_{\ell_2}^{\Delta \Delta} \left( z_3, z_2 \right) } \, .
\ee
We set $z_1=z_2=z_3=1$ and show the dependence on the three
multipoles. Considering equal redshifts has the advantage of
  allowing a straightforward relation between multipoles and modes in
  Fourier space $k \approx \ell / r(z)$, given the line-of-sight
  comoving distance $r(z)$. The choice $\ell_1=200$ then
  corresponds to a comoving wavenumber $k_1 \approx 0.06/$Mpc (or
  equivalently a 100 Mpc comoving scale), which is quite linear at
    $z=1$. The leading terms are typically of the same order,
$\sim 0.1$ to $0.2$ for the different configurations. The density
dipole term amounts to about 10 to 30\% of the monopole while the
quadrupole is 3 to 10\%. The terms shown in the second row of
fig.~\ref{Fig_den_other} are significantly smaller. Indeed the latter,
even if computed at the same redshift, contain integrals of spherical
Bessel function which are out of phase leading to a strong suppression
of their amplitudes.
All velocity terms have the largest power in the squeezed limit
(top-left corner in the figures), whereas the density monopole somewhat prefers the equilateral shape.
Note, however, that the bispectra are nearly constant with little
variation over the range shown in the figures.

\begin{figure}[t]
\begin{center}
\includegraphics[width=0.45\textwidth]{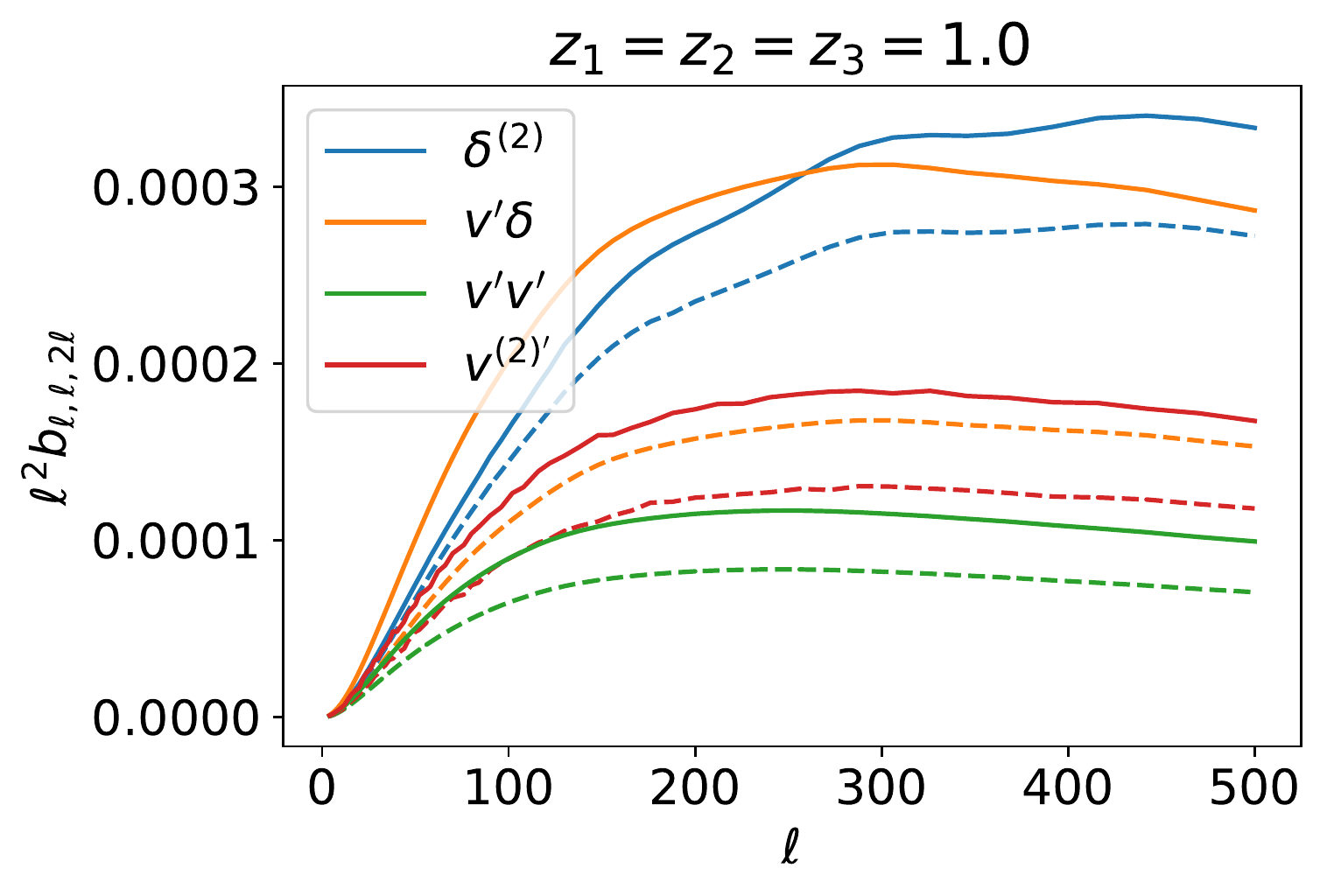}
\includegraphics[width=0.45\textwidth]{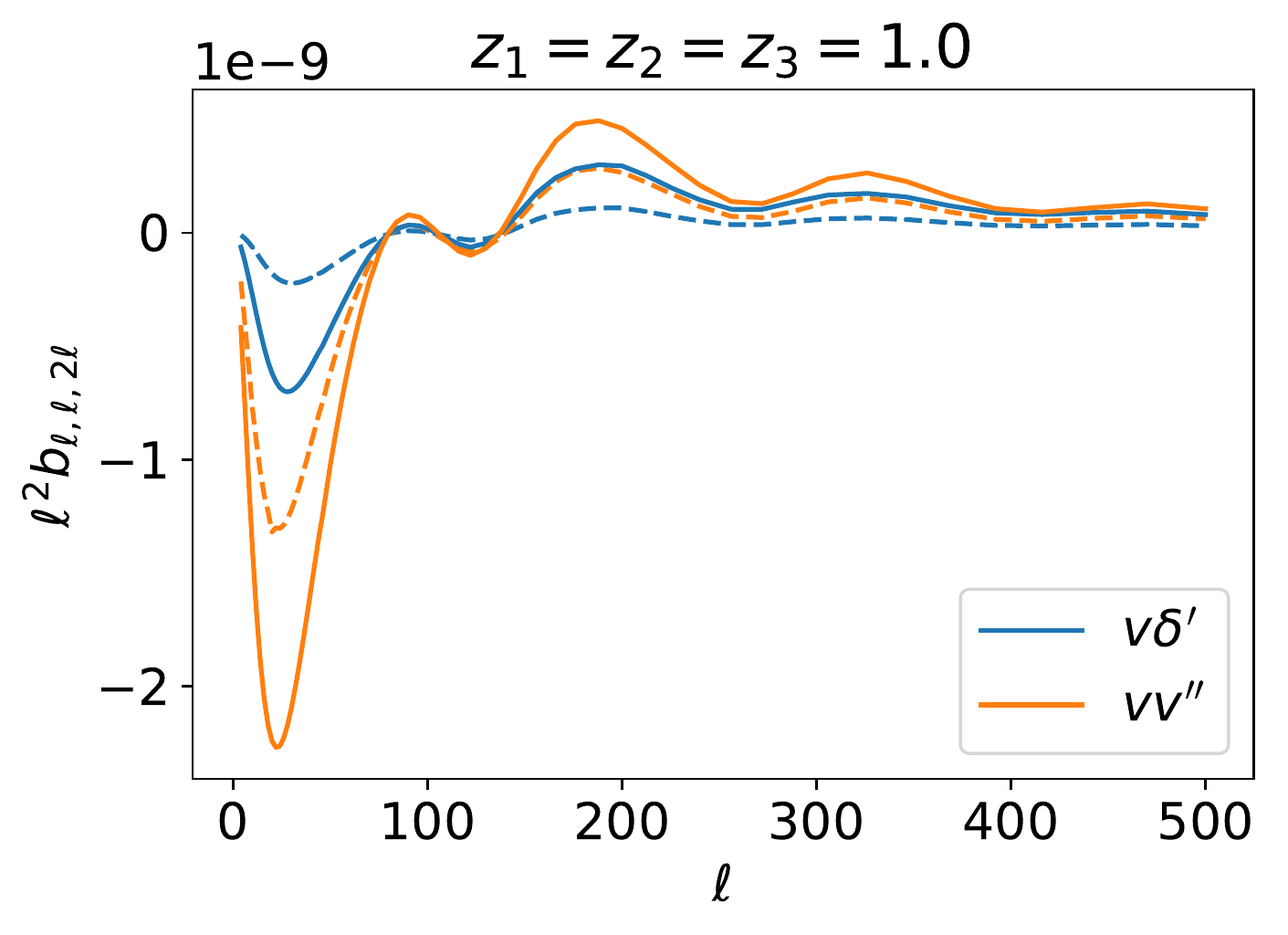}
\caption{Comparison of the reduced bispectrum contributions discussed
  in Section~\ref{sec:angbisp} as a function of
  $\ell\equiv\ell_1=\ell_2=\ell_3/2$. Dashed lines show the unbiased
  bispectra, $b_1=1$, $b_2=b_s=0$. For clarity we split the main and
  subdominant contributions for this configuration into the left and
  right panels, respectively.}
\label{fig:b_contr}
\end{center}
\end{figure}

In figure \ref{fig:b_contr} we show the different contributions to the
tree-level bispectrum in the configuration $\ell_1=\ell_2=\ell_3/2$ at
equal redshifts $z_1=z_2=z_3=1$ as a function of $\ell_1$. Clustering
bias (figure \ref{f:bias}) is very important and due to a negative
  $b_2(z)$ the term $b^{ v' \delta}_{\ell_1 \ell_2 \ell_3}$
  dominates  at large scales. Consistently with the previous plots,
contributions involving the velocity terms
$b^{ \delta' v}_{\ell_1 \ell_2 \ell_3}$ and
$b^{ v'' v}_{\ell_1 \ell_2 \ell_3}$ are significantly suppressed
with respect to the other terms. The physical oscillations in these
velocity terms are roughly in phase.

\begin{figure}[t]
\begin{center}
\includegraphics[width=0.7\textwidth]{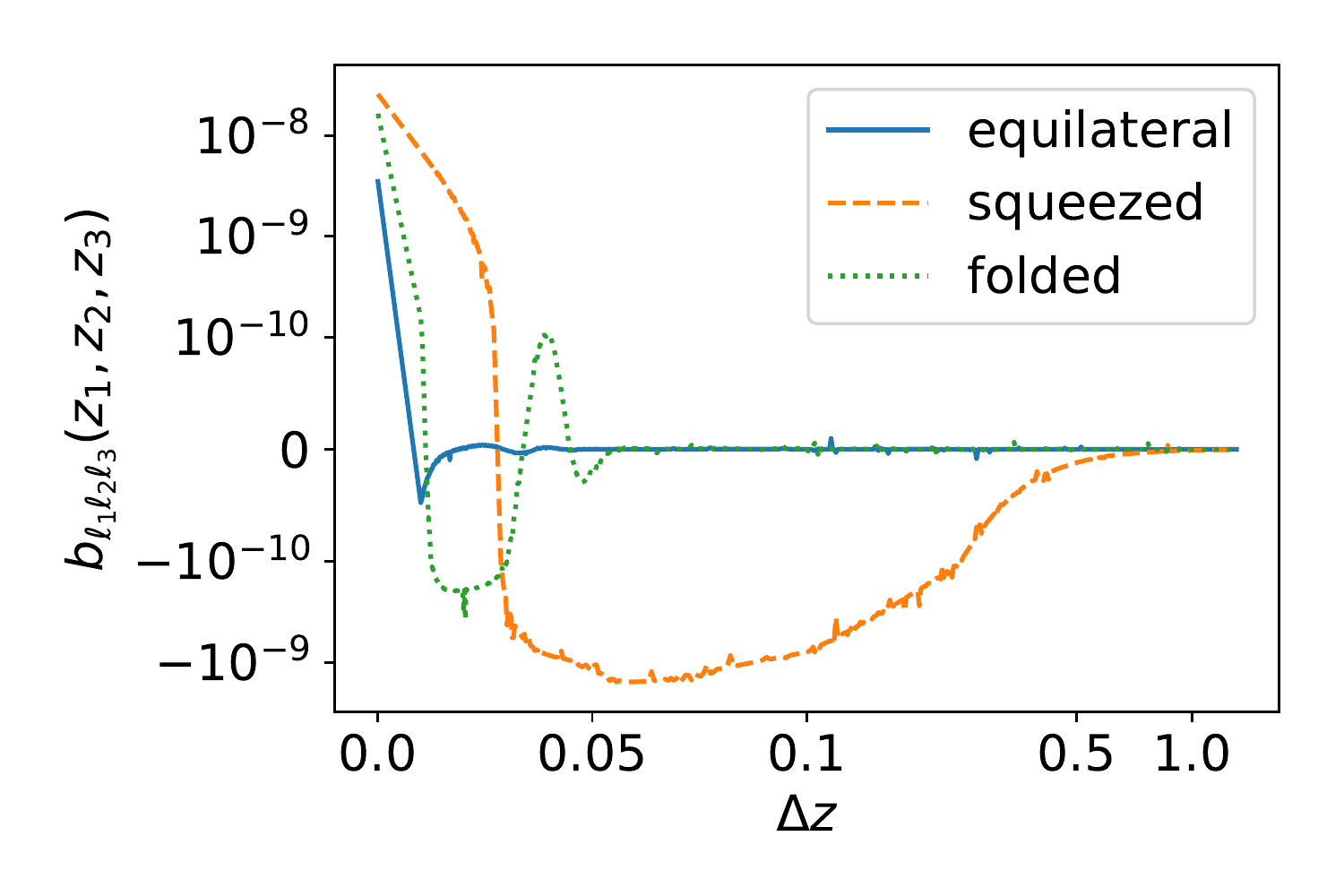}
\caption{Density reduced bispectrum as a function of $\Delta z$, where
  $z_1=0.7$ and $z_2=z_3=z_1+\Delta z$, and for different multipole
  configurations: equilateral ($\ell_1=\ell_2=\ell_3=400$), squeezed
  ($\ell_1=4$, $\ell_2=\ell_3=400$) and folded
  ($\ell_1=\ell_2=\ell_3/2=200$). Small-scale noise is due to
    numerical errors, negligible for our purposes.}
\label{fig:dens_neq_z}
\end{center}
\end{figure}

In figure~\ref{fig:dens_neq_z} we consider the density term as a
  function of redshift difference for $z_1=0.7$ and $z_2=z_3=z_1 +\De z$
  in the following configurations\footnote{We
    refer to a given shape considering only multipole triangles. Of
    course, a given shape at equal or at different redshifts
    corresponds to a different configuration in comoving space. In
    particular, care should be taken when comparing to results
    obtained in Fourier space (at equal redshifts).} (always such that
  $\ell_1+\ell_2+\ell_3$ is even):

\begin{itemize}
\item Equilateral, $\ell_1=\ell_2=\ell_3$.
\item Squeezed, $\ell_1 \ll \ell_2=\ell_3$.
\item Folded, $\ell_1=\ell_2=\ell_3/2$.
\end{itemize}

The density bispectrum is  suppressed by several orders of
  magnitude already at relatively small redshift differences $\Delta z$.  In all cases the
  bispectrum reaches a negative minimum and then tends to
  zero. In the
  squeezed case the trough is wider and the suppression is
  somewhat less severe.

We also compare the density bispectrum to its cosmic variance. We
use eq.~(\ref{eq:gb_red}) to write the angle-averaged bispectrum as
\begin{equation}
  B_{\ell_1\ell_2\ell_3} =
  \sqrt{\frac{(2\ell_1+1)(2\ell_2+1)(2\ell_3+1)}{4\pi}}
  \tj{\ell_1}{\ell_2}{\ell_3}{0}{0}{0} b_{\ell_1\ell_2\ell_3} \;,
\end{equation}
and we define the signal-to-noise ratio for a fixed multipole configuration as
\begin{equation} \label{eq:sn}
  \frac{S}{N} =
  \frac{|B_{\ell_1\ell_2\ell_3}|}{\sigma_{B_{\ell_1\ell_2\ell_3}}} \;.
\end{equation}
The variance $\sigma_{B_{\ell_1\ell_2\ell_3}}$ is calculated in Appendix~\ref{sec:cv}.
Correlations at different redshifts are subdominant compared to equal
redshift contributions in the computation of the cosmic variance
$\sigma_{B_{\ell_1\ell_2\ell_3}}^2$ given by equation
(\ref{eq:B_covarince_sumleven}). Hence, we neglect terms
$c_{\ell}(z_i, z_j)$ with $z_i \neq z_j$ in the variance.

\begin{figure}[t]
\begin{center}
\includegraphics[width=0.49\textwidth]{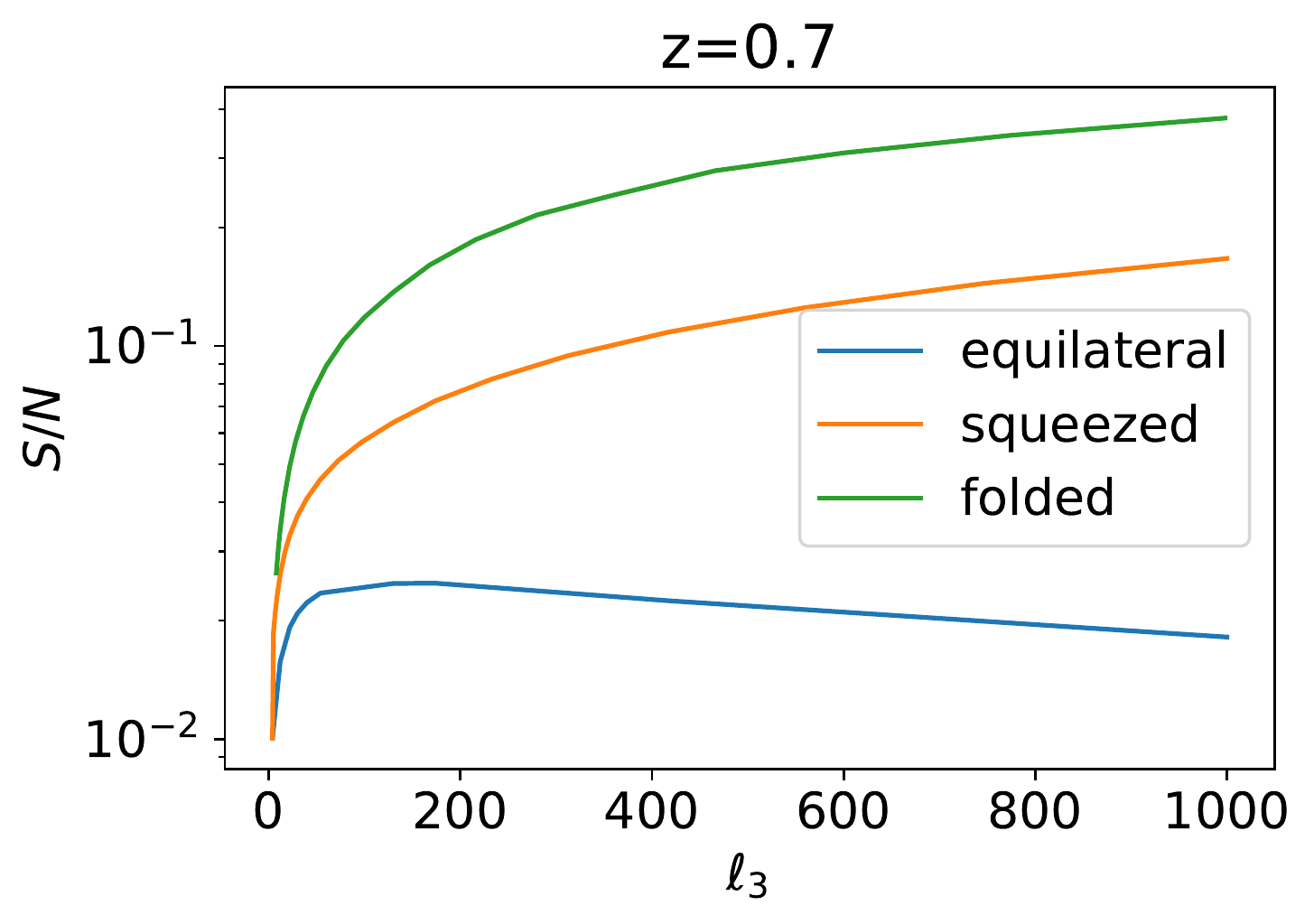}
\includegraphics[width=0.49\textwidth]{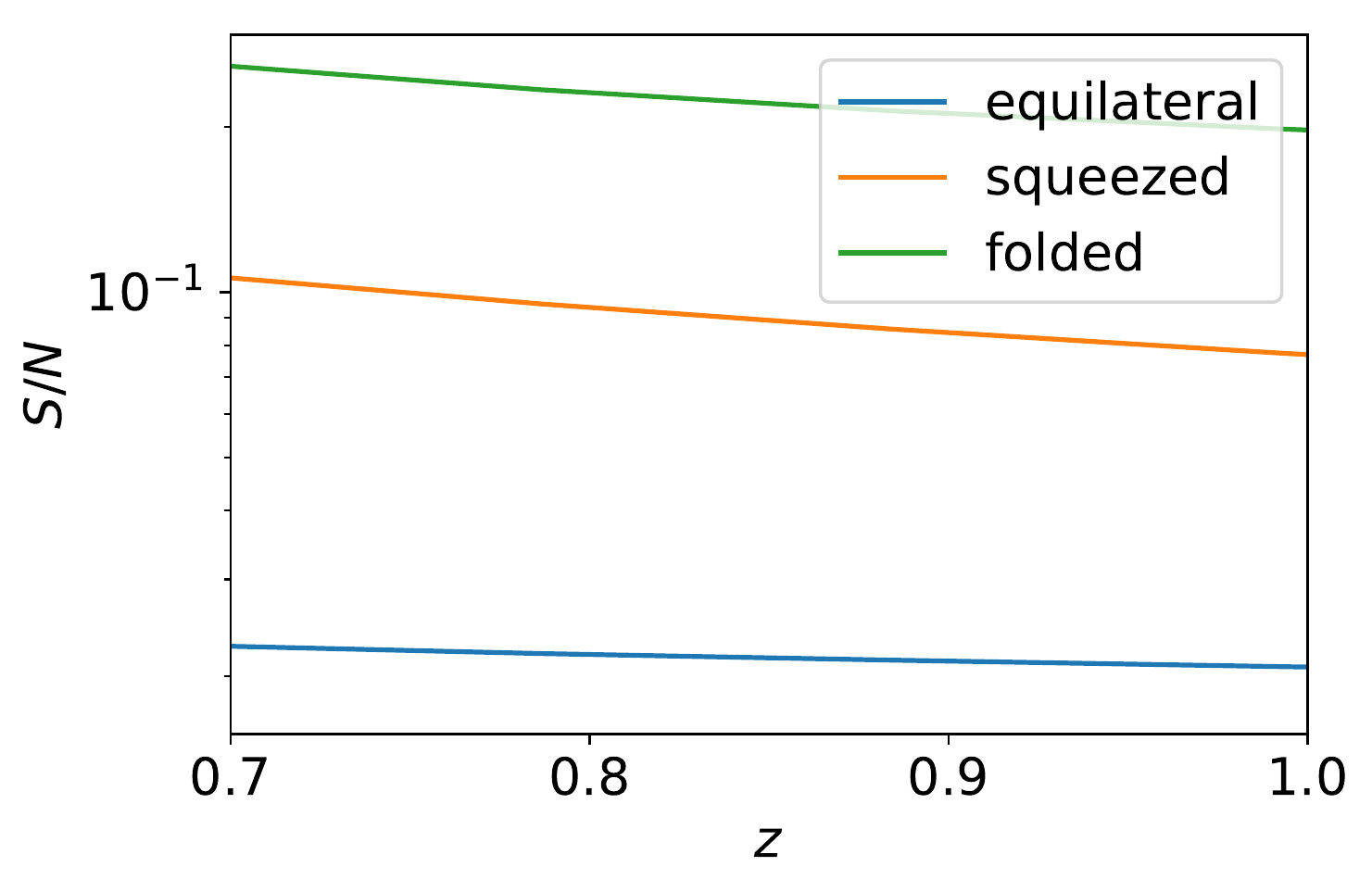}
\caption{Signal-to-noise (cosmic variance) ratio for the density
  bispectrum.  \emph{Left panel:} dependence on the largest multipole,
  $\ell_3$, for different configurations at $z_1=z_2=z_3=z=0.7$. In the squeezed
  configuration we consider $\ell_1=4$. \emph{Right panel:} dependence
  on redshift for the equilateral ($\ell_1=\ell_2=\ell_3=400$),
  squeezed ($\ell_1=4$, $\ell_2=\ell_3=400$) and folded
  ($\ell_1=\ell_2=\ell_3/2=200$) shapes.}
\label{fig:sn}
\end{center}
\end{figure}

For all shapes\footnote{While strictly speaking the squeezed limit
    corresponds to $\ell_1\ll\ell_2=\ell_3$, here we loosen this
    definition.}  we find a $S/N \gtrsim 0.01$ at $z_1=z_2=z_3=0.7$,
  leading to a presumably large cumulative $S/N$ when summing over all
  triangles. The results also show that, compared to cosmic variance,
  the density contribution is more sensitive to the folded and
  squeezed cases and the equilateral case is more strongly
  suppressed. We checked that the folded configuration also yields
  better results than the case $\ell_1\neq\ell_2\neq\ell_3$ (not shown
  in the figure).  In the squeezed case, we verified that increasing the
  long mode $\ell_1$ decreases the $S/N$. Note the turnover in $S/N$
  in the equilateral configuration, due to the fact that the signal
  decreases faster than the variance at large multipoles. In all cases
  the $S/N$ decreases with redshift, as non-Gaussianities get
  weaker. However, for different redshifts with $\Delta z \gtrsim 0.1$, for high redshifts, $z>2$ and wide redshift bins, lensing may not be
  negligible and hence should be included both in the signal and in
  the variance; see Ref.~\cite{DiDio:2015bua} for numerical results including lensing.

We stress that the $S/N$ estimates presented in this section are only
meant to study the effect of cosmic variance on different bispectrum
configurations. A realistic estimate must include the effect of finite
radial selection functions and of shot-noise, the latter being
particularly relevant at small scales.\footnote{The validity of a
  tree-level bispectrum should also be better assessed at the largest
  multipoles and smallest redshifts here considered.}

%%%%%%%%%%%%%%%%%%%%%%%%%%%%%%
\subsection{Limber approximation for density and RSD \label{s:Limber}}
\begin{figure}[t]
\begin{center}
\includegraphics[width=0.44\textwidth]{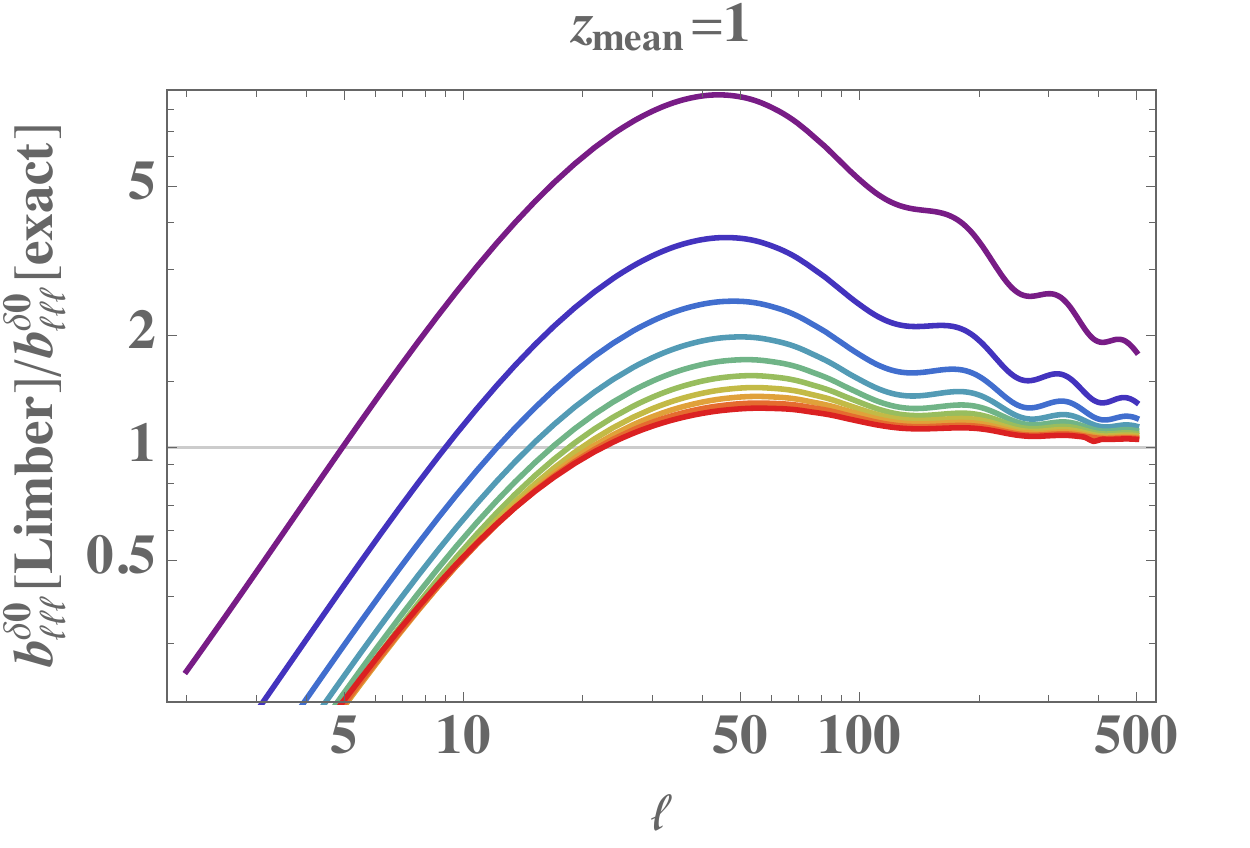}
\includegraphics[width=0.08\textwidth]{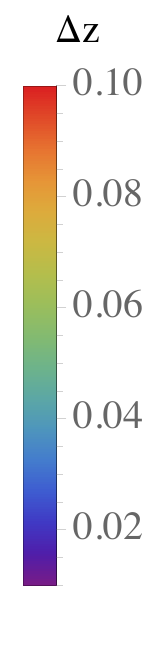}
\includegraphics[width=0.44\textwidth]{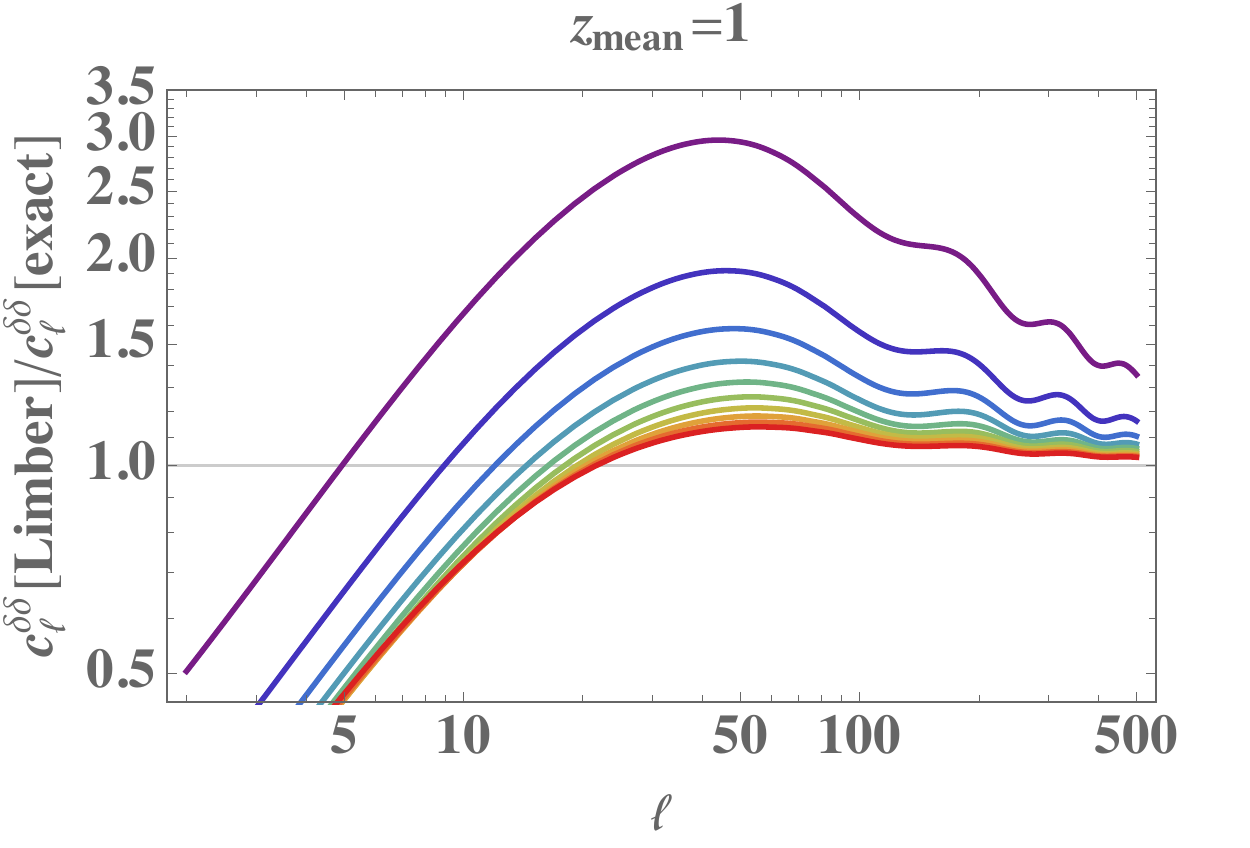}
\caption{On the left panel we plot the ratio between the Limber
  approximation and the exact solution of the monopole of the density
  contribution to the bispectrum in the equilateral
  configuration. Colors indicate different window sizes, from
  full-width $\Delta z =0.01$ (violet) to $\Delta z =0.1$ (red). On
  the right panel we plot the analogous ratio in terms of binned
  angular power spectra, showing that the Limber approximation is much
  less accurate for the bispectrum for z-bins of the same width. }
\label{fig:limber_density}
\end{center}
\end{figure}

In this section we introduce a redshift binning. To reduce the numerical effort we adopt the Limber approximation~\cite{Limber:1954zz,LoVerde:2008re}
\be
\label{limber}
\frac{2}{\pi} \int dk k^2 f( k ) j_\ell \left( k r_1 \right) j_\ell \left( k r_2 \right) \simeq f \left( \frac{\ell+1/2}{r_1}\right) \frac{\delta_D \left( r_1 - r_2 \right)  }{r_1^2} \, .
\ee
For the density contribution we consider the monopole of the reduced bispectrum and we compare the Limber approximation with the exact solution. We also neglect RSD contributions to the linear perturbations. Introducing redshift binning with a normalized window function centered at $z$, $W(z,z')$, we define the following z-binned bispectrum (denoted with an over-bar)
\be
\bar b_{\ell_1 \ell_2 \ell_3 } \left( z_1 , z_2 , z_3 \right) = \int dz'_1 dz'_2 dz'_3 W (z_1,z'_1) W (z_2,z'_2)W (z_3,z'_3) b_{\ell_1 \ell_2 \ell_3 } \left( z'_1 , z'_2 , z'_3 \right) \, .
\ee
For the density bispectrum we apply the Limber approximation directly on eq.~\eqref{7}
\bea
 && \bar b^{\delta^{(2)}}_{\ell_1\ell_2 \ell_3}(z_1,z_2,z_3) ~=~\frac{16}{\pi^3} \int dz'_1 dz'_2 dz'_3 W_1 (z'_1) W_2 (z'_2)W_3 (z'_3) \int dk_1 dk_2 dk_3 \int d\chi \chi^2 k_1^2 k_2^2
k_3^2
\nonumber \\
&& T_\delta \left( k_2, z'_1\right) T_\delta \left( k_3, z'_1\right)
T_\delta \left( k_2, z'_2\right) T_\delta \left( k_3, z'_3\right) P_R
\left( k_2 \right) P_R\left( k_3 \right) F_2 \left( k_1 , k_2 , k_3
\right)
\nonumber \\
&& j_{\ell_1 } \left( k_1 r'_1 \right) j_{\ell_2 } \left( k_2 r'_2
\right) j_{\ell_3 } \left( k_3 r'_3 \right) j_{\ell_1 } \left( k_1 \chi
\right) j_{\ell_2 } \left( k_2 \chi \right) j_{\ell_3 } \left( k_3
  \chi \right)
+ \circlearrowleft
\nonumber \\
&&
\simeq~ 2 \int \frac{d\chi}{\chi^4} \frac{W(z_1,z(\chi))W(z_2, z(\chi))W(z_3,z(\chi))}{\left( dr/dz |_{r = \chi} \right)^3} B_{\delta\delta\delta}\left( \bar k_1, \bar k_2, \bar k_3\right) \left( D_1 \left( z\left( \chi \right) \right) \right)^4 + \circlearrowleft \\
&& \mbox{where~~}  \bar k_i= \frac{\ell_i+1/2}{\chi}\,. \nonumber
\eea
Here we have used  the same binning for each variable
$z_1, z_2, z_3$ and $B_{\delta\delta\delta} \left( k_1, k_2, k_3
\right) $ is the density bispectrum in Fourier space, see e.g.~Ref.~\cite{Bernardeau:2001qr}.
In fig.~\ref{fig:limber_density} we show the accuracy of Limber approximation for different z-binning. It is evident that the Limber approximation is not accurate to describe redshift bins more narrow than about~$\Delta z \sim 0.1$ for $\ell<500$ if we require an accuracy of 10\% or better.

For the RSD term we can not apply Limber approximation on all the Bessel functions, due to its second derivative appearing in eq.~\eqref{bisp_RSD2}.
Therefore, we consider
\be \label{bessel_2ndDer}
j''_\ell \left( x \right) = \frac{\left(\ell^2-\ell-x^2\right) j_{\ell}(x)}{x^2}+ \frac{2 j_{\ell+1}(x)}{x}
\ee
such that we can apply the Limber approximation on the first term directly. For the second term we generalize the Limber approximation, which is derived from the approximation
\be
j_\ell \left( x \right) \simeq \sqrt{\frac{\pi}{2 \ell +1}} \delta_D \left( \ell + \frac{1}{2} - x \right) \, ,
\ee
yielding
\be
\label{limber_generalize}
\frac{2}{\pi} \int dk k^2 f( k ) j_{\ell} \left( k r_1 \right) j_{\ell+1} \left( k r_2 \right) \simeq
\sqrt{\frac{2 \ell +1 }{2 \ell + 3 }} \ f \left( \frac{\ell+1/2}{r_1}\right) \frac{\delta_D \left( r_1 \frac{2 \ell +3 }{2 \ell +1}- r_2 \right)  }{r_1^2} \, .
\ee
Using the ordinary Limber approximation~\eqref{limber} on the first term of eq.~\eqref{bessel_2ndDer} and the generalized approximation~\eqref{limber_generalize} on the second term, we obtain
\bea
&& \bar b^{v^{(2)'}}_{\ell_1 \ell_2 \ell_3 }(z_1,z_2,z_3)
= -\frac{16}{\pi^3} \int  dz'_1  dz'_2  dz'_3 W(z_1, z'_1) W(z_2, z'_2 )  W(z_3,z'_3)
f(z'_1) b_1(z'_2) b_1(z'_3)\times
\nonumber \\
&&
\int dk_1 dk_2 d k_3 k_1^2 k_2^2 k_3^2
G_2 \left( k_1, k_2, k_3 \right) P_R \left( k_2 \right) P_R( k_3) \times
\nonumber \\
&&
j''_{\ell_1} \left( k_1 r'_1 \right) j_{\ell_2} \left( k_2 r'_2 \right)
j_{\ell_3} \left( k_3 r'_3 \right) T_\delta \left( k_2, z'_1 \right)
T_\delta\left( k_3, z'_1 \right) T_\delta \left( k_2, z'_2 \right)
T_\delta \left( k_3 , z'_3 \right)
\nonumber \\
&&
 \times\int_0^\infty d\chi \chi^2 j_{\ell_1} \left( k_1 \chi
\right)j_{\ell_2} \left( k_2 \chi \right)j_{\ell_3} \left( k_3 \chi
\right) + \circlearrowleft
\nonumber \\
\simeq~ &&
 2
\frac{1+ 8 \ell_1}{\left( 1+2 \ell_1 \right)^2}
 \!\!\int\! \frac{d \chi}{\chi^4}\frac{ W(z_1,z(\chi))W(z_2, z(\chi))W(z_3,z(\chi))}{ \left(dr/dz  |_{r= \chi}\right)^3} f \left(\chi \right) b_1 \left( \chi \right)^2 G_2 \left(  \bar k_1, \bar k_2,\ \bar k_3 \right)\times
\nonumber \\
&&
P_R \left(  \bar k_1\right)P_R \left(  \bar k_3\right) T_\delta \left( \bar k_2 , \chi \right)^2 T_\delta \left(  \bar k_3 , \chi \right)^2
\nonumber \\
&&
- \frac{8 }{ \sqrt{  2 \ell_1 +1} \sqrt{  2 \ell_1 +3} } \int \frac{d\chi}{\chi^4}  \frac{ W \left(z_2,z(\chi) \right)W \left(z_3,z(\chi) \right)}{\left(dr/dz  |_{r= \chi}\right)^2} \frac{ W \left(z_1, z\left(\frac{2 \ell_1 +3 }{2 \ell_1 +1}\chi \right) \right)}{dr/dz  |_{r= \frac{2 \ell_1 +3 }{2 \ell_1 +1}\chi}}
\times  \nonumber \\
&&
 f \left(  \frac{2 \ell_1 +3 }{2 \ell_1 +1}\chi \right) b_1 \left( \chi \right)^2G_2 \left(  \bar k_1, \bar k_2, \bar k_3 \right) P_R \left(  \bar k_2 \right)P_R \left( \bar k_3\right)
\times \nonumber \\
&&
 T_\delta \left(  \bar k_2 , \chi \right) T_\delta \left( \bar k_2, \frac{2 \ell +3 }{2 \ell +1}\chi \right)  T_\delta \left( \bar k_3 , \chi \right) T_\delta \left(  \bar k_3, \frac{2 \ell_1 +3 }{2 \ell_1 +1}\chi \right)
  ~ + ~   \circlearrowleft \, .
 \nonumber \\
 &&  \mbox{where~~}  \bar k_i= \frac{\ell_i+1/2}{\chi}\,.
\eea

\begin{figure}[t]
\begin{center}
\includegraphics[width=0.45\textwidth]{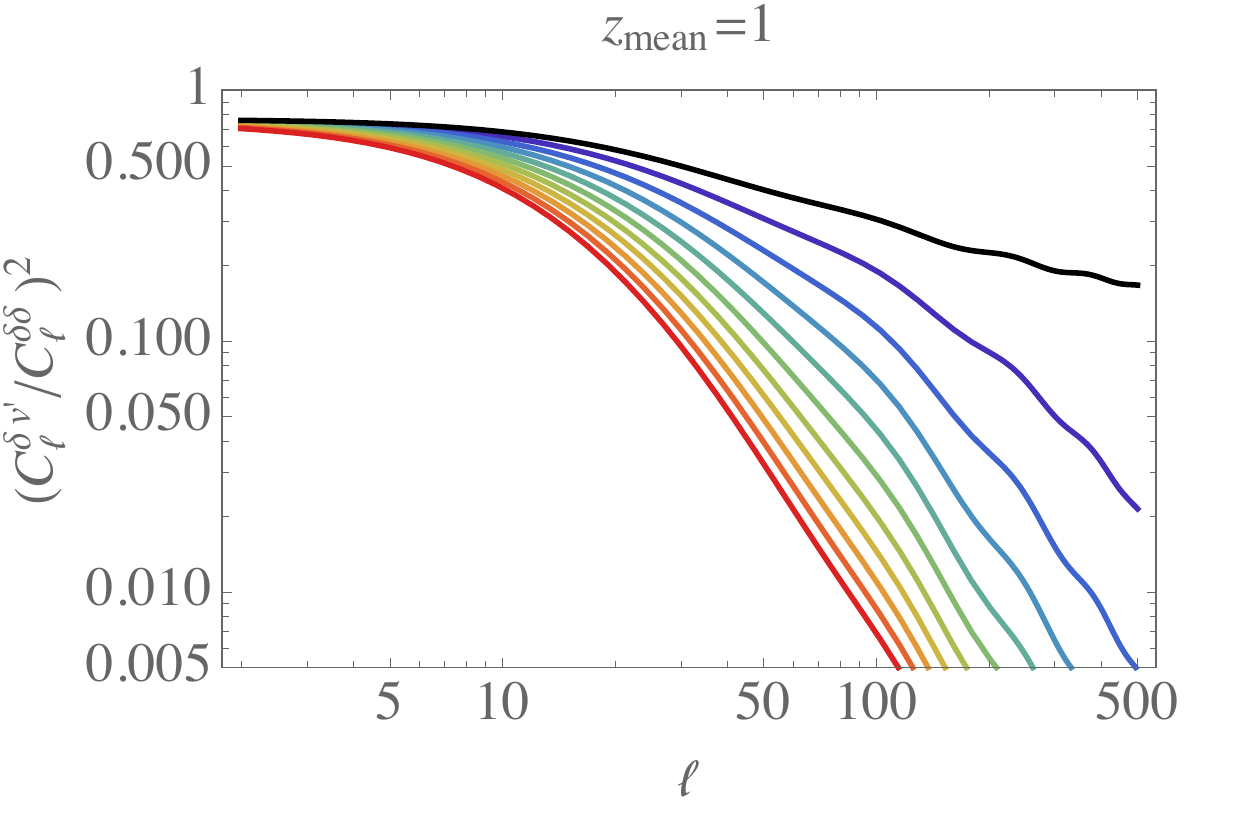}
\includegraphics[width=0.07\textwidth]{figs/legend.pdf}
\includegraphics[width=0.45\textwidth]{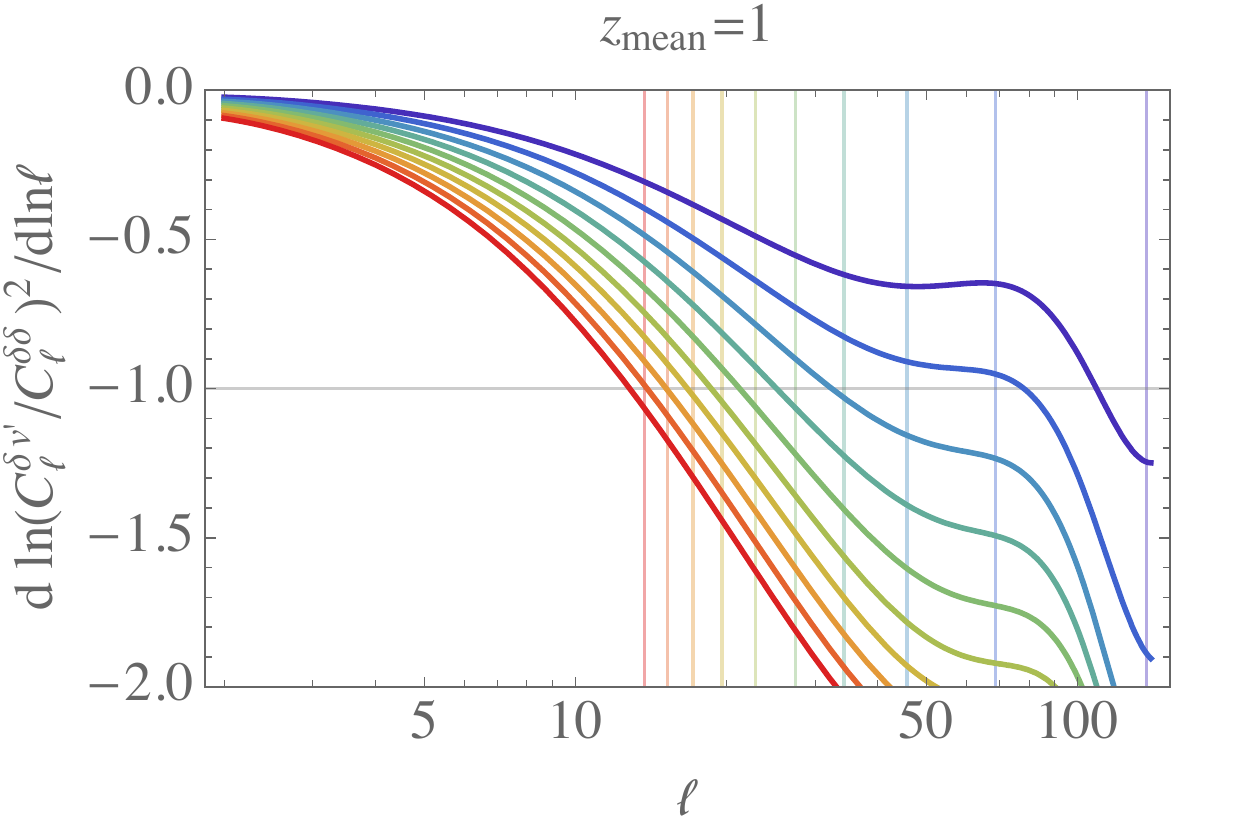}
\caption{Comparison of RSD vs density power spectra at different z-bin
  for the power spectrum. At large $\ell$ they scale as
  $\ell^{-1}$. The left panel is showing the square of this ratio.
  Here we use it as a proxy for the suppression of RSD with respect to
  the density in the bispectrum as function of the size of window
  function. In the right panel we show the slope of the ratio between
  the square of RSD and density spectra. The vertical lines denote
  $\ell_{*}= r(z_m)/(r(z_m+\Delta z/2)-r(z_m-\Delta z/2)) $. This
  scale roughly determines the change of slope from $-1$ to $-2$. In
  other words, for a given $\Delta z$ at scales smaller than
  $\ell_{*}$ RSD is strongly suppressed. On the other hand, for
  $\ell<\ell_*$ Limber approximation for the RSD contribution is not
  reliable.}
\label{fig:cl_limber}
\end{center}
\end{figure}

Interestingly the scale dependence of the two contributions at large $\ell$ (limit of validity of Limber approximation) are given respectively by
\be
  \frac{1+ 8 \ell_1}{\left( 1+2 \ell_1 \right)^2} = \frac{2}{\ell_1} - \frac{7}{4 \ell_1^2} + \mathcal{O} \left( \ell_1^{-3} \right)
 \ee
 and
 \be
 \frac{4}{ \sqrt{  2 \ell_1 +1} \sqrt{  2 \ell_1 +3} } = \frac{2}{\ell_1} - \frac{2}{\ell_1^2} + \mathcal{O} \left( \ell_1^{-3} \right) \, .
\ee
The two leading contributions, that scale as $\ell_1^{-1}$, cancel exactly leaving the z-binned RSD bispectrum scaling as $\ell^{-2}$ with respect to the density perturbations.
Considering that the contribution of RSD to the z-binned power spectrum scales as $\ell^{-1}$ with respect to density perturbations and that we do expect the bispectrum to scale roughly like the square of the power spectrum, we can use the latter as a proxy to investigate the importance of RSD for different z-binning. This finding, together with the corresponding results for the power spectrum shown in Fig.~\ref{fig:cl_limber}, suggests that for binned bispectra with a wide enough binning compatible with the Limber approximation, RSD is negligibly small. Only for very slim redshift bins, as  are possible for spectroscopic surveys, is RSD measurable. Furthermore, for such bins the Limber approximation cannot be trusted.

%%%%%%%%%%%%%%%%%%%%%%%%%%%%
\section{The HI intensity mapping angular bispectrum}
\label{sec:hi_im}

Intensity mapping experiments provide maps of the emitted intensity of a given molecular or atomic line. Here we consider the $21$cm line from the spin flip (hyperfine structure) of neutral hydrogen (HI). We are interested in the low redshift, i.e.~the post-reonization universe, on scales where it is well described by quasi-linear physics~\cite{Wyithe:2007gz,Wyithe:2008mv}.
A $21$cm intensity survey does not resolve single galaxies, but it measures the superposed diffuse emission of several sources. Therefore intensity mapping experiments do typically not provide high angular resolution. However, as they  directly measure frequencies, they allow for very narrow redshift bins.
Moreover, $21$cm intensity surveys face the challenge to clean the cosmological signal of foregrounds
 which are  several orders of magnitude larger.
Foreground cleaning approaches~\cite{Alonso:2014dhk} are based on the different frequency dependence of foregrounds which are typically smooth while the 21cm line is very sharply peaked. This allows us to recover the 21cm intensity very well on small radial scales, but recovery becomes increasingly difficult on large scales, on which also the 21 cm intensity should have smooth correlations. This may limit for practical purposes the use of HI mapping tomography since the long range radial correlations are lost by the cleaning of the much stronger foreground.
We plan in the future to include the angular formalism developed in this work in a full 3-dimensional Fourier angular spectrum~\cite{Heavens:1994iq,Rassat:2011aa,Yoo:2013tc} to precisely quantify and possibly overcome this limitation.

On the other hand, the HI mapping angular bispectrum, being limited to very thin z-bins, requires one to include RSD developed in our formalism. Indeed, while in galaxy surveys broad z-bins can strongly reduce the impact of RSD on the observable power spectra or higher order statistics, broad z-bins in HI mapping remain foreground dominated.
Previous works, for instance~\cite{Schmit:2018rtf}, have not included RSD in the bispectrum analysis. Moreover they based their results on the Limber approximation, which, as we have shown, fails completely for the slim z-bins required by HI mapping.

Without the need to resolve single galaxies, intensity mapping experiments can more efficiently cover a larger fraction of the sky and go deeper in redshift, providing more information on the largest scales we can observe.
A $21$cm intensity mapping survey provides the brightness temperature $T_b\left( \bn , z \right)$ of neutral hydrogen~\cite{Furlanetto:2006jb,Wyithe:2007gz} as a function of sky direction $\bn$ and observed redshift $z$. We can therefore define the brightness temperature fluctuation as
\be
\Delta_T\left( \bn , z \right) = \frac{T_b\left( \bn , z \right) - \langle T_b\rangle \left( z \right)}{\langle T_b\rangle \left( z \right)}
\ee
where $\langle (\cdots) \rangle$ denotes the angular average at fixed observed redshift $z$.
 The brightness
temperature in 21cm intensity is  proportional to the observed neutral
hydrogen number density $n_\text{HI}$ per unit
  surface and per redshift bin~(see \cite{Hall:2012wd}, eq.~(22))
\be\label{temp_gal_rel}
T_b \left( \bn , z \right) \propto \frac{{n_\text{HI}} \left( \bn , z \right)}{d_A^2 \left( \bn , z \right)}\,,
\ee
where
$d_A$ is the angular diameter distance.

Therefore, to any order,  the HI intensity mapping fluctuation differs only by a convergence term (lensing) from the galaxy number counts. Since in our work we do not consider any term induced by the lensing potential (see Ref.~\cite{Jalivand:2018vfz} for HI intensity mapping lensing at second and third order) our results can be directly interpreted also in terms of the HI angular bispectrum, providing we use the appropriate bias parameters~\cite{Umeh:2015gza}.

%%%%%%%%%%%%%%%%%%%%%%%%%%%%%%%%%%%%%%%%%%%%%%%%%
\section{Conclusions}
\label{sec:conclusions}

In this work we have computed the tree level angular-redshift bispectrum for number counts taking into account the contribution from density and redshift space distortions. For equal redshifts, $z_1=z_2=z_3$ and very narrow redshift bins as are achieved for spectroscopic number count surveys and for intensity mapping, these are the only relevant contributions for $\ell_i \gtrsim 10$.  On large scales, $\ell\sim10$ relativistic and wide angle effects are relevant, while for unequal redshifts and for wide redshift bins, lensing terms are relevant.

We have presented results using the bias prescription expected to hold for a Euclid-like survey. We have found that the normalized bispectrum contributions are relatively constant and of the order 0.1 to 0.2. All contributions are very similar and not strongly scale or shape dependent, apart from the $\delta'v$ and $v''v$ terms which are suppressed by several orders of magnitude due to the fact that these terms are out of phase.

The $S/N$ of the folded configuration is largest, about a factor of 3 larger than the squeezed contribution, and up to a
factor of 30 larger than the equilateral contribution where the
$S/N$ of a single mode remains of order 0.01 up to $\ell=1000$, while the single mode values for the folded and squeezed configurations at $\ell\sim 1000$ become of order 0.1. In this analysis only cosmic variance is included in the noise, so it certainly over-estimates the value, but it indicates that it should be straightforward to detect this bispectrum from nonlinearities by summing over several modes.

We have considered redshift binning and shown that the Limber approximation is unreliable if the redshift bins are too slim. For $\Delta z=0.01$ we find deviations from the correct density bispectrum of up to a factor 8.  This is due to the relatively high UV power of density perturbations which are a factor $k^2/\HH^2$ larger than the gravitational potential. For sufficiently wide redshift bins which average over scales smaller than $\Delta z/\HH(z)$, this UV power is reduced which renders the Limber approximation less inaccurate. Even though we cannot compute the binned RSD without Limber approximation we are convinced that the same is true for RSD since it has the same UV behavior as the density term. Introducing significant redshift binning reduces the RSD bispectrum by about a factor $\ell^{-2}$ faster than the density bispectrum so that in the regime where the Limber approximation is reasonably reliable, $\ell\gtrsim100$ and $\Delta z\gtrsim 0.1$, RSD can actually be neglected.

The decomposition into spherical harmonics has always been used for
the statistical description and analysis of CMB physics, but the
subtle UV issues we investigated here do not affect CMB observables
because the latter are proportional to the metric perturbations and
not to their second derivatives, i.e., density and RSD. Therefore,
their transfer functions decay much faster with $k$ and one does not
encounter UV divergences.

The present study of the bispectrum is still preliminary and can be improved and probably sped up in multiple ways.
For example, while in this work we express the bispectrum for all the
terms except RSD as products of generalized angular power spectra, in the future we
plan to replace numerical line-of-sight integrals by the more efficient
FFTlog approaches advocated in~\cite{Assassi:2017lea}. This has proven
to speed-up computations by a factor of order $\mathcal{O}(100)$
\cite{Schoneberg:2018fis}. This numerical improvement is especially
important when  using redshift binning which increases the number of integrals
and when we want to be efficient enough
 to sample cosmological parameters with MCMC methods based on
a bispectrum likelihood.

%%%%%%%%%%%%%%%%%%%%%%%%%%%%%%%%%%%%%%%%%%%%%%%%%
\section*{Acknowledgements}
It is a pleasure to thank Simone Ferraro for useful discussions.
ED is supported by the Swiss National Science Foundation (No.~171494).
RD is supported by the Swiss National Science Foundation.
FM is supported by the Research Project FPA2015-68048-C3-3-P
[MINECO-FEDER], and the Centro de Excelencia Severo Ochoa Program
SEV-2016-0597. RM and OU are supported by the UK STFC grant ST/N000668/1. RM is also supported by the South African SKA Project and the National Research Foundation (Grant No. 75415).
%%%%%%%%%%%%%%%%%%%%%%%%%%%%%%%%%%%%%%%%%%%%%%%%%
\appendix

%%%%%%%%%%%%%%%%%%%%%%%%%%%%%%%%%
\section{Details for the RSD bispectrum}
\label{app:rsd}

The aim of this section is to provide the bispectrum for RSD,
described in eq.~\eqref{v2_eq}, in terms of products of 1-dimensional
integrals, useful for numerical computations. The main task is to apply the derivative to the products of Bessel functions (last two lines of eq.~\eqref{v2_eq}) and to rewrite them in terms of spherical Bessel functions of different order.
We warn the reader, that due to the recursive relation for spherical Bessel functions, the final results can be rewritten in several  other equivalent forms.

Without the integrated terms along $\chi$ we can rewrite eq.~\eqref{v2_eq} in terms of the generalized spectra~\eqref{eq:c_ll}, as follows
\bea
&&b^{v^{(2)'}}_{\ell_1 \ell_2 \ell_3 } \left( z_1 , z_2 , z_3 \right)
\supset 2 f(z_1) b_1(z_2) b_1(z_3)
\times
\nonumber \\
&&
\Bigg\{
\frac{1}{224} \left(\frac{16 {\ell_3}^2}{4 {\ell_2} ({\ell_2}+1)-3}+\frac{4 (8 {\ell_1} ({\ell_1}+1)+88 {\ell_2} ({\ell_2}+1)-115)}{4 {\ell_2} ({\ell_2}+1)-3}+\frac{4 {\ell_2}^2+8 {\ell_1} ({\ell_1}+1)-49}{2 {\ell_3}-1}
\right.
\nonumber \\
&&
\left.
+\frac{-4 {\ell_2}^2-8 {\ell_1} ({\ell_1}+1)+49}{2 {\ell_3}+3}\right) {c_{\ell_2}}({z_2},{z_1}) {c_{\ell_3}}({z_3},{z_1})
\nonumber \\
&&
%%%%%%%
+\frac{1}{56} \left(-\frac{2 {\ell_3}^2}{4 {\ell_2} ({\ell_2}+2)+3}+\frac{2 {\ell_3}}{2 {\ell_2}+1}+\frac{-4 {\ell_1} ({\ell_1}+1)+2 {\ell_2} (11 {\ell_2}+9)+2}{4 {\ell_2} ({\ell_2}+2)+3}+\frac{({\ell_1}-2) ({\ell_1}+3)}{2 {\ell_3}-1}
\right.
\nonumber \\
&&
\left.
-\frac{({\ell_1}-2) ({\ell_1}+3)}{2 {\ell_3}+3}\right) {{c_{\ell_2 \ell_2+2 }}}({z_2},{z_1}) {c_{\ell_3}}({z_3},{z_1})
\nonumber \\
&&
%%%%%%%
+\left(-4 {\ell_3}^4-8 {\ell_2} {\ell_3}^3+(-8 {\ell_1} ({\ell_1}+1)+44 {\ell_2} ({\ell_2}+1)+19) {\ell_3}^2+2 ({\ell_2} (22 {\ell_2}+29)
\right.
\nonumber \\
&&
\left.
-4 {\ell_1} ({\ell_1}+1)) {\ell_3}+9 {\ell_3}-3 {\ell_2} (27 {\ell_2}+13)+4 {\ell_1} ({\ell_1}+1) \left(2 {\ell_2}^2+1\right)+3\right)
\nonumber \\
&&
\times
\frac{{{c_{\ell_2 \ell_2 -2 }}}({z_2},{z_1}) {c_{\ell_3}}({z_3},{z_1})}{28 \left(4 {\ell_2}^2-1\right) (4 {\ell_3} ({\ell_3}+1)-3)}
\nonumber \\
&&
%%%%%%%
+\frac{1}{224} \left(\frac{8 {\ell_1} ({\ell_1}+1)+4 ({\ell_2}-2) {\ell_2}+3}{2 {\ell_3}+1}+\frac{2 (8 {\ell_1} ({\ell_1}+1)+44 {\ell_2} ({\ell_2}+1)-81)}{4 {\ell_2} ({\ell_2}+1)-3}
\right.
\nonumber \\
&&
\left.
+\frac{-4 {\ell_2}^2-8 {\ell_1} ({\ell_1}+1)+49}{2 {\ell_3}-1}\right) {c_{\ell_2}}({z_2},{z_1}) {{c_{\ell_3 \ell_3- 2 }}}({z_3},{z_1})
\nonumber \\
&&
%%%%%%%
+\left(\left(12 (2 {\ell_3}+1)^2-8 {\ell_1} ({\ell_1}+1)\right) {\ell_2}^2-4 (2 {\ell_1} ({\ell_1}+1)-6 {\ell_3}-3) (2 {\ell_3}-1) {\ell_2}
\right.
\nonumber \\
&&
\left.
-4 (2 {\ell_1} ({\ell_1}+1)-3) ({\ell_3}-1) {\ell_3}-9\right)\frac{ {{c_{\ell_2 \ell_2 -2 }}}({z_2},{z_1}) {{c_{\ell_3 \ell_3- 2 }}}({z_3},{z_1})}{56 \left(4 {\ell_2}^2-1\right) \left(4 {\ell_3}^2-1\right)}
\nonumber \\
&&
%%%%%%%
+\left(-8 ({\ell_2}-{\ell_3}+1) ({\ell_2}-{\ell_3}+2) {\ell_1}^2-8 ({\ell_2}-{\ell_3}+1) ({\ell_2}-{\ell_3}+2) {\ell_1}
\right.
\nonumber \\
&&
\left.
+3 (2 {\ell_2}+1) (2 {\ell_3}+1) (2 {\ell_3}+{\ell_2} (4 {\ell_3}+2)+5)\right)\frac{ {{c_{\ell_2 \ell_2+2 }}}({z_2},{z_1}) {{c_{\ell_3 \ell_3- 2 }}}({z_3},{z_1})}{56 (4 {\ell_2} ({\ell_2}+2)+3) \left(4 {\ell_3}^2-1\right)}
\nonumber \\
&&
%%%%%%%
+\frac{1}{224} \left(-\frac{8 {\ell_1} ({\ell_1}+1)+4 ({\ell_2}-2) {\ell_2}+3}{2 {\ell_3}+1}+\frac{2 (8 {\ell_1} ({\ell_1}+1)+44 {\ell_2} ({\ell_2}+1)-81)}{4 {\ell_2} ({\ell_2}+1)-3}
\right.
\nonumber \\
&&
\left.
+\frac{4 {\ell_2}^2+8 {\ell_1} ({\ell_1}+1)-49}{2 {\ell_3}+3}\right) {c_{\ell_2}}({z_2},{z_1}) {{c_{\ell_3 \ell_3 +2 }}}({z_3},{z_1})
\nonumber \\
&&
%%%%%%%
+\left(-8 ({\ell_2}-{\ell_3}-2) ({\ell_2}-{\ell_3}-1) {\ell_1}^2-8 ({\ell_2}-{\ell_3}-2) ({\ell_2}-{\ell_3}-1) {\ell_1}
\right.
\nonumber \\
&&
\left.
+3 (2 {\ell_2}+1) (2 {\ell_3}+1) (2 {\ell_3}+{\ell_2} (4 {\ell_3}+2)+5)\right)\frac{ {{c_{\ell_2 \ell_2 -2 }}}({z_2},{z_1}) {{c_{\ell_3 \ell_3 +2 }}}({z_3},{z_1})}{56 \left(4 {\ell_2}^2-1\right) (4 {\ell_3} ({\ell_3}+2)+3)}
\nonumber \\
&&
%%%%%%%
+\left(-8 ({\ell_2}+{\ell_3}+2) ({\ell_2}+{\ell_3}+3) {\ell_1}^2-8 ({\ell_2}+{\ell_3}+2) ({\ell_2}+{\ell_3}+3) {\ell_1}
\right.
\nonumber \\
&&
\left.
+3 (2 {\ell_2}+1) (2 {\ell_3}+1) (2 {\ell_3}+{\ell_2} (4 {\ell_3}+2)-3)\right)\frac{ {{c_{\ell_2 \ell_2+2 }}}({z_2},{z_1}) {{c_{\ell_3 \ell_3 +2 }}}({z_3},{z_1})}{56 (2 {\ell_2}+1) (2 {\ell_2}+3) (2 {\ell_3}+1) (2 {\ell_3}+3)}
\nonumber \\
&&
%%%%%%%
+\frac{(10 {\ell_2}+24 {\ell_3}+19) {\ ^{-1}{c_{\ell_3 \ell_3 -1 }}}({z_3},{z_1}) {\ ^{1}{c_{\ell_2 \ell_2 -1 }}}({z_2},{z_1})}{224 {\ell_3}+112}
\nonumber \\
&&
%%%%%%%
+\frac{(-10 {\ell_2}+24 {\ell_3}+5) {\ ^{-1}{c_{\ell_3 \ell_3 +1 }}}({z_3},{z_1}) {\ ^{1}{c_{\ell_2 \ell_2 -1 }}}({z_2},{z_1})}{224 {\ell_3}+112}
\nonumber \\
&&
%%%%%%%
+\frac{(2 {\ell_2}+8 {\ell_3}+3) {\ ^{-1}{c_{\ell_3 \ell_3 -1 }}}({z_3},{z_1}) {\ ^{1}{c_{\ell_2 \ell_2 -3 }}}({z_2},{z_1})}{224 {\ell_3}+112}
\nonumber \\
&&
%%%%%%%
+\frac{(-2 {\ell_2}+8 {\ell_3}+5) {\ ^{-1}{c_{\ell_3 \ell_3 +1 }}}({z_3},{z_1}) {\ ^{1}{c_{\ell_2 \ell_2 -3 }}}({z_2},{z_1})}{224 {\ell_3}+112}
\nonumber \\
&&
%%%%%%%
+\frac{(-10 {\ell_2}+16 {\ell_3}+17) {\ ^{-1}{c_{\ell_3 \ell_3 -1 }}}({z_3},{z_1}) {\ ^{1}{c_{\ell_2 \ell_2 +1 }}}({z_2},{z_1})}{224 {\ell_3}+112}
\nonumber \\
&&
%%%%%%%
+\frac{(10 {\ell_2}+16 {\ell_3}-1) {\ ^{-1}{c_{\ell_3 \ell_3 +1 }}}({z_3},{z_1}) {\ ^{1}{c_{\ell_2 \ell_2 +1 }}}({z_2},{z_1})}{112 (2 {\ell_3}+1)}
\nonumber \\
&&
%%%%%%%
+\frac{(-2 {\ell_2}+8 {\ell_3}+1) {\ ^{-1}{c_{\ell_3 \ell_3 -1 }}}({z_3},{z_1}) {\ ^{1}{c_{\ell_2 \ell_2 +3 }}}({z_2},{z_1})}{224 {\ell_3}+112}
\nonumber \\
&&
%%%%%%%
+\frac{(2 {\ell_2}+8 {\ell_3}+7) {\ ^{-1}{c_{\ell_3 \ell_3 +1 }}}({z_3},{z_1}) {\ ^{1}{c_{\ell_2 \ell_2 +3 }}}({z_2},{z_1})}{224 {\ell_3}+112}
\nonumber \\
&&
%%%%%%%
+\frac{(24 {\ell_2}+10 {\ell_3}+19) {\ ^{-1}{c_{\ell_2 \ell_2 -1 }}}({z_2},{z_1}) {\ ^{1}{c_{\ell_3 \ell_3 -1 }}}({z_3},{z_1})}{224 {\ell_2}+112}
\nonumber \\
&&
%%%%%%%
+\frac{(24 {\ell_2}-10 {\ell_3}+5) {\ ^{-1}{c_{\ell_2 \ell_2 +1 }}}({z_2},{z_1}) {\ ^{1}{c_{\ell_3 \ell_3 -1 }}}({z_3},{z_1})}{224 {\ell_2}+112}
\nonumber \\
&&
%%%%%%%
+\frac{(8 {\ell_2}+2 {\ell_3}+3) {\ ^{-1}{c_{\ell_2 \ell_2 -1 }}}({z_2},{z_1}) {\ ^{1}{c_{\ell_3 \ell_3 -3 }}}({z_3},{z_1})}{224 {\ell_2}+112}
\nonumber \\
&&
%%%%%%%
+\frac{(8 {\ell_2}-2 {\ell_3}+5) {\ ^{-1}{c_{\ell_2 \ell_2 +1 }}}({z_2},{z_1}) {\ ^{1}{c_{\ell_3 \ell_3 -3 }}}({z_3},{z_1})}{224 {\ell_2}+112}
\nonumber \\
&&
%%%%%%%
+\frac{(16 {\ell_2}-10 {\ell_3}+17) {\ ^{-1}{c_{\ell_2 \ell_2 -1 }}}({z_2},{z_1}) {\ ^{1}{c_{\ell_3 \ell_3 +1 }}}({z_3},{z_1})}{224 {\ell_2}+112}
\nonumber \\
&&
%%%%%%%
+\frac{(16 {\ell_2}+10 {\ell_3}-1) {\ ^{-1}{c_{\ell_2 \ell_2 +1 }}}({z_2},{z_1}) {\ ^{1}{c_{\ell_3 \ell_3 +1 }}}({z_3},{z_1})}{112 (2 {\ell_2}+1)}
\nonumber \\
&&
%%%%%%%
+\frac{(8 {\ell_2}-2 {\ell_3}+1) {\ ^{-1}{c_{\ell_2 \ell_2 -1 }}}({z_2},{z_1}) {\ ^{1}{c_{\ell_3 \ell_3 +3 }}}({z_3},{z_1})}{224 {\ell_2}+112}
\nonumber \\
&&
%%%%%%%
+\frac{(8 {\ell_2}+2 {\ell_3}+7) {\ ^{-1}{c_{\ell_2 \ell_2 +1 }}}({z_2},{z_1}) {\ ^{1}{c_{\ell_3 \ell_3 +3 }}}({z_3},{z_1})}{224 {\ell_2}+112}
\Bigg\} \, .
\label{eqA1}
\eea

The full RSD bispectrum is therefore given by eq.~\eqref{eqA1}
plus the $\chi$-integral given below for completeness,
\bea \label{eq:int-chi-term}
&&- 2\left( 4\pi \right)^2 f(z_1)
b_1(z_2) b_1(z_3) \int \frac{dk_2}{k_2} \frac{dk_3}{k_3} \mathcal{P}_R
\left( k_2 \right) \mathcal{P}_R\left( k_3 \right)
\nonumber \\
&& T_\delta \left( k_2, z_1\right) T_\delta \left( k_3, z_1\right)
T_\delta \left( k_2, z_2\right) T_\delta \left( k_3, z_3\right)
j_{\ell_2 } \left( k_2 r_2 \right) j_{\ell_3 } \left( k_3 r_3
\right)\frac{1}{2\ell_1 + 1}
\nonumber \\
&& \int d\chi \left( \ell_1 \left( \ell_1 -1\right)
  \frac{r_1^{\ell_1-2}}{\chi^{\ell_1-1}} \Theta_H \left( \chi - r_1
  \right) + \left( \ell_1 +1 \right) \left( \ell_1 +2 \right)
  \frac{\chi^{\ell_1+2}}{r_1^{3+\ell_1}} \Theta_H \left( r_1 -\chi
  \right) \right)
\nonumber \\
&& j_{\ell_2 } \left( k_2 \chi \right) j_{\ell_3 }\left( k_3 \chi
\right) G^{(0)}_2\left( k_2, k_3 \right) \, .
\eea
We solve eq.~(\ref{eq:int-chi-term}) numerically as a 3-dimensional integral
using the Suave Monte Carlo algorithm available from the Cuba library
\cite{Hahn:2004fe}.\footnote{\url{http://www.feynarts.de/cuba/}. The
  Suave algorithm does not need to evaluate the integrand at the
  integration boundaries, so the semi-infinite range of integration
  can be simply mapped to the unitary interval through a change of
  coordinates. Alternatively, note that the $\chi$-integrand is
  strongly peaked around $\chi \sim r_1$, so the semi-infinite range
  can be safely restricted to, e.g., $\chi \in [r_1/10,
  10\,r_1]$.}
\begin{figure}[t]
\begin{center}
\includegraphics[width=0.49\textwidth]{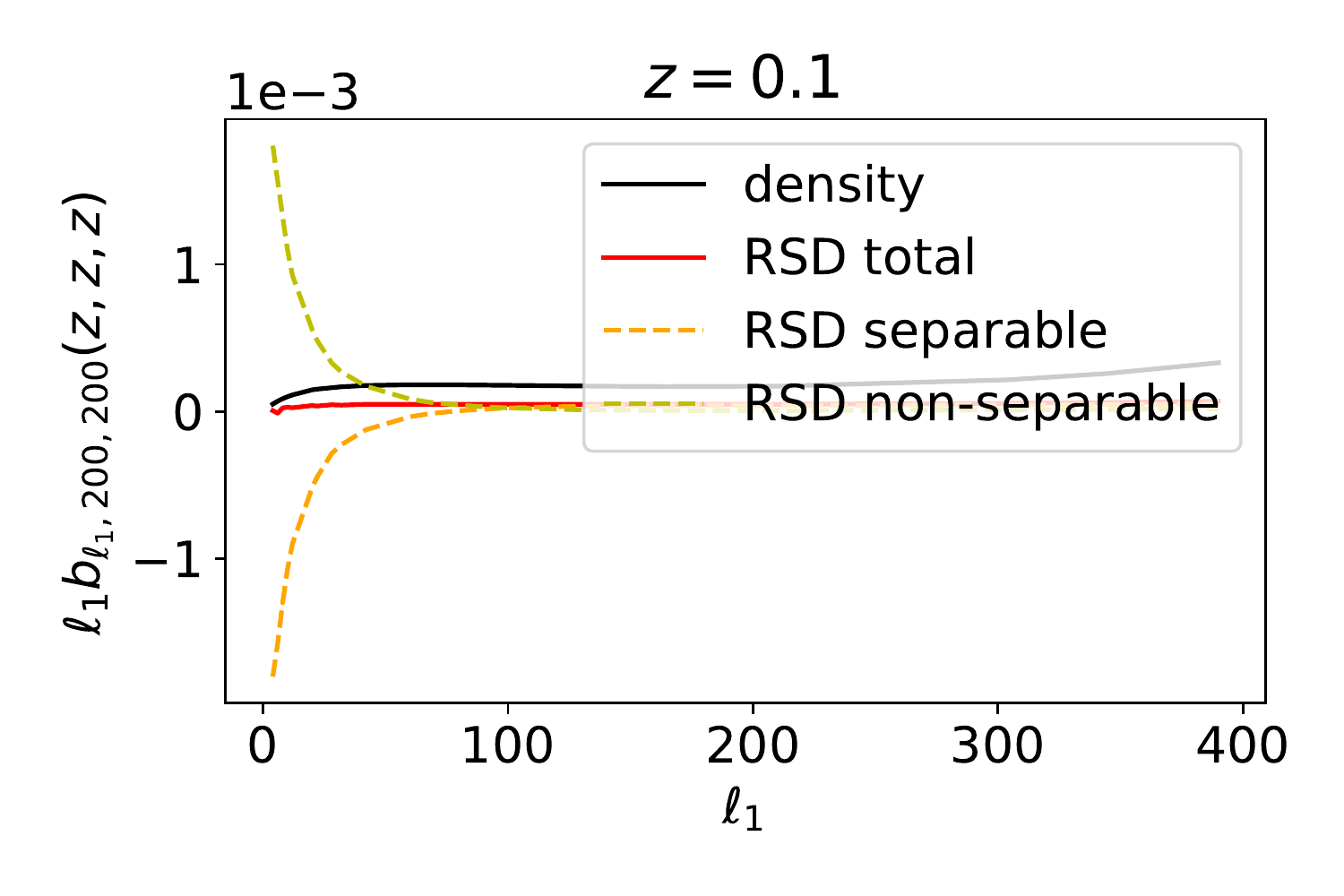}
\includegraphics[width=0.49\textwidth]{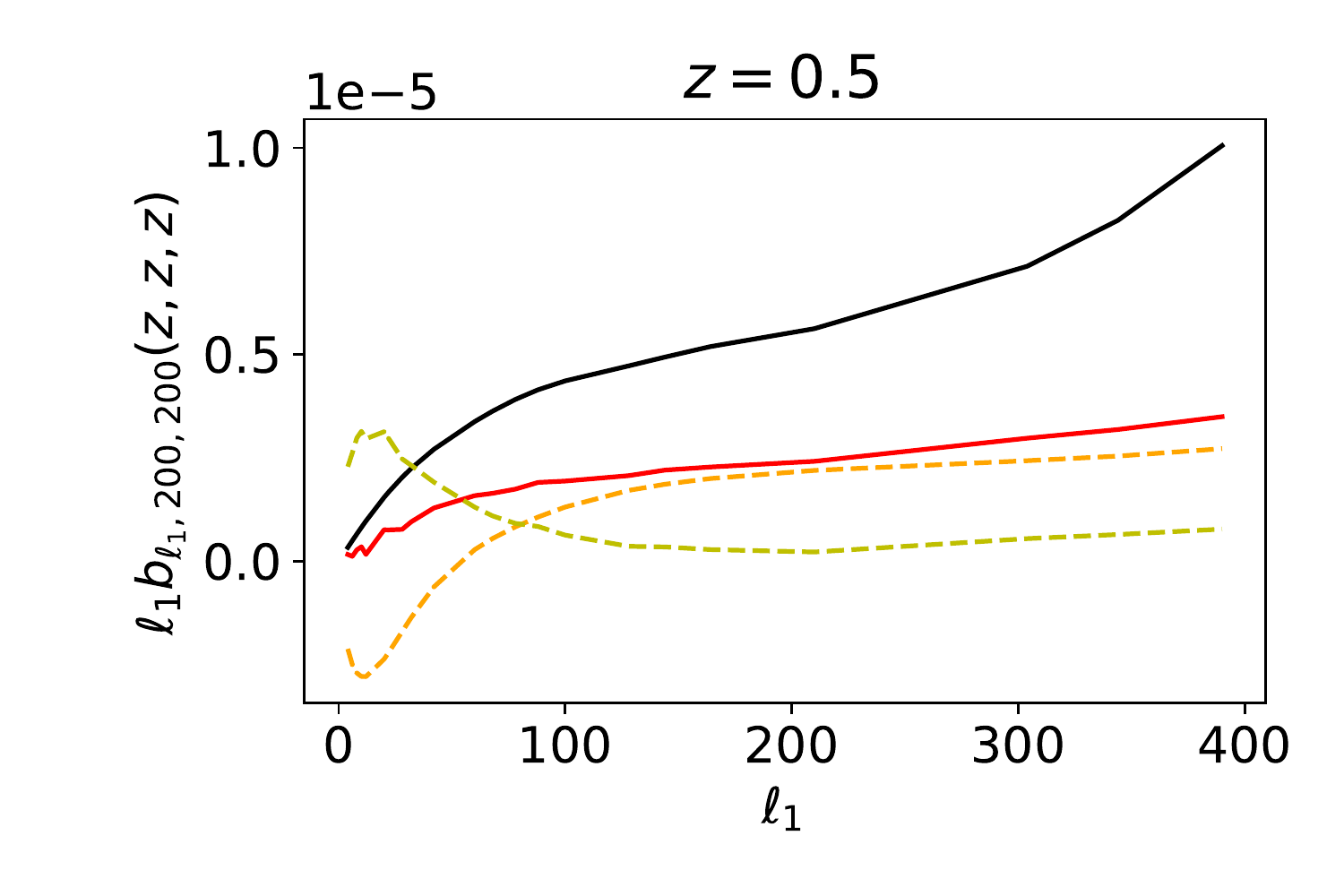}
\caption{Cancellation between separable, eq.~(\ref{eqA1}), and
  non-separable, eq.~(\ref{eq:int-chi-term}), RSD reduced bispectrum
  terms. For comparison we plot also the density contribution. We
  consider equal redshifts ($z=0.1$ on the left panel, and $z=0.5$ on
  the right) and $\ell_2=\ell_3=200$. Cosmological parameters are the
  same as in section \ref{sec:numerical}, but for a simpler
  interpretation we do not include clustering bias nor linear RSD.}
\label{fig:sep_nonsep}
\end{center}
\end{figure}

Figure \ref{fig:sep_nonsep} shows a striking cancellation between the
separable RSD reduced bispectrum term, eq.~(\ref{eqA1}), and the
non-separable one, eq.~(\ref{eq:int-chi-term}), up to two orders of
magnitude at the smallest multipoles at $z=0.1$. This suggests that a
more convenient factorization where the sum is carried out
analytically is desirable. Also note that the non-separable term
becomes sub-leading at small scales ($\ell_1 \gtrsim 100$ in the
plotted configurations).

%%%%%%%%%%%%%
\section{Density $\delta^{(2)}$: Alternative derivation}
\label{sec:b_dens_app}
We give an alternative derivation of the density bispectrum to that
presented in section~\ref{sec:b_dens}, following the computation
presented in section~\ref{sec:b_rsd} for RSD. This is useful to
validate the method as we compared the analytical and numerical
results for the density reduced bispectrum obtained in the two
approaches. Here we neglect clustering bias for brevity, but it is
clear how to re-introduce it.

Given eqs.~(\ref{eq:dens_corr}), (\ref{eq:dens2_expr}), we obtain the
3-point function \bea \label{eq:bispdens_dkg} \langle .. \rangle &=&
\frac{2}{\left( 2 \pi \right)^6} \int d^3k_1 d^3k_2 d^3k_3 \delta_D
\left( \bk_1 + \bk_2 +\bk_3 \right)T_\delta \left( k_2, z_1 \right)
T_\delta \left( k_3, z_1 \right) T_\delta \left( k_2, z_2 \right)
T_\delta \left( k_3, z_3 \right)
\nonumber \\
&& \qquad \qquad \times F_2 \left( k_1 ,k_2, k_3 \right) P_R \left( k_2
\right) P_R \left( k_3 \right) e^{i \left( \bk_1 \cdot \bn_1 r_1+\bk_2
    \cdot \bn_2 r_2+\bk_3 \cdot \bn_3 r_3\right)} \,.  \eea
We expand the Fourier modes as \be e^{i \bk \cdot \bn r} = 4 \pi
\sum_{\ell m} i^\ell j_\ell \left( k r \right) Y_{\ell m } \left( \bn
\right) Y^*_{\ell m } \left( {\hat \bk} \right) \ee
and the Dirac-delta distribution as
\bea \label{eq:deltaD3} &&\delta^{(3)}_D
\left( \bk_1 + \bk_2 + \bk_3 \right) = \int \frac{d^3 x}{\left( 2 \pi
  \right)^3} e^{i \left( \bk_1 + \bk_2 + \bk_3 \right) \cdot {\bf x}}
\nonumber \\
&=& 8 \int d\chi \chi^2 d \Omega_\bn \prod_{p=1}^3 \left\{
  \sum_{\ell'_p m'_p} \left( - i \right)^{\ell'_p} j_{\ell'} \left(
    k_p \chi\right) Y_{\ell'_p m'_p} \left( \hat \bk_p \right)
  Y^*_{\ell'_p m'_p} \left( \bn \right) \right\} \,.\eea
Integrating now over the angular part of the Fourier integrals we obtain
\be \langle .. \rangle = \sum_{\ell_1 \ell_2 \ell_3 m_1 m_2 m_3}
B^{m_1 m_2 m_3 }_{\ell_1 \ell_2 \ell_3} Y_{\ell_1 m_1} \left( \bn_1
\right) Y_{\ell_2 m_2} \left( \bn_2 \right) Y_{\ell_3 m_3} \left(
  \bn_3 \right) \ee with \bea B^{m_1 m_2 m_3 }_{\ell_1 \ell_2
  \ell_3}&=& \Gaunt{m_1}{m_2}{m_3}{\ell_1}{\ell_2}{\ell_3} b_{\ell_1
  \ell_2 \ell_3}\\
  b_{\ell_1\ell_2 \ell_3}&=&\frac{16}{\pi^3} \int dk_1 dk_2 dk_3 \int d\chi \chi^2 k_1^2 k_2^2
k_3^2
\nonumber \\
&& T_\delta \left( k_2, z_1\right) T_\delta \left( k_3, z_1\right)
T_\delta \left( k_2, z_2\right) T_\delta \left( k_3, z_3\right) P_R
\left( k_2 \right) P_R\left( k_3 \right) F_2 \left( k_1 , k_2 , k_3
\right)
\nonumber \\
&& j_{\ell_1 } \left( k_1 r_1 \right) j_{\ell_2 } \left( k_2 r_2
\right) j_{\ell_3 } \left( k_3 r_3 \right) j_{\ell_1 } \left( k_1 \chi
\right) j_{\ell_2 } \left( k_2 \chi \right) j_{\ell_3 } \left( k_3
  \chi \right)
\label{7}
\eea
where
\bea
 F_2 \left( \bk_2 , \bk_3 \right) &=& \frac{5}{7} + \frac{1}{2} \frac{\bk_2 \cdot \bk_3 }{k_2 k_3} \left( \frac{k_2}{k_3} + \frac{k_3}{k_2} \right) + \frac{2}{7} \left( \frac{\bk_2 \cdot \bk_3}{k_2 k_3} \right)^2
 \\
 F_2 \left( k_1 , k_2 ,k_3 \right) &=&\frac{5}{7} + \frac{1}{4} \frac{k_1^2 - k_2^2 - k_3^2}{k_2 k_3} \left( \frac{k_2}{k_3} + \frac{k_3}{k_2} \right) + \frac{1}{14} \left( \frac{k_1^2 - k_2^2 - k_3^2}{k_2 k_3} \right)^2
 \nonumber \\
 &=&
 -\frac{5 \left(k_2^2-k_3^2\right)^2}{28 k_2^2 k_3^2}
 +\frac{3}{28} \left(\frac{1}{k_2^2}+\frac{1}{k_3^2}\right) k_1^2
 +\frac{k_1^4}{14 k_2^2 k_3^2}
  \nonumber \\
 &\equiv&
 F_2^{(0)} \left( k_2 , k_3 \right) + F_2^{(2)} \left( k_2 , k_3 \right) k_1^2+ F_2^{(4)} \left( k_2 , k_3 \right) k_1^4 \, .
\eea
By expanding the integral~\eqref{7}, we note that it depends on $k_1$ only through
\bea
\int dk_1 k_1^2 j_{\ell_1}  \left( k_1 r_1 \right)   j_{\ell_1}\left( k_1 \chi \right) &=& \frac{\pi}{2 \chi^2} \delta_D \left( \chi - r_1 \right) \, ,
\\
\int dk_1 k_1^4 j_{\ell_1}  \left( k_1 r_1 \right)   j_{\ell_1}\left( k_1 \chi \right) &=&
 \frac{\pi}{2  r_{ 1}^2}  \left[ - \frac{\partial^2}{\partial \chi^2} - \frac{2}{\chi} \frac{\partial}{\partial \chi} + \frac{\ell_1 \left( \ell_1 +1 \right) }{\chi^2} \right] \delta_D \left( \chi - r_1 \right)\, ,
 \label{intk4}
 \\
 \int dk_1 k_1^6 j_{\ell_1} \left( k_1 r_1 \right) j_{\ell_1} \left(
   k_1 \chi \right) &=& \frac{\pi}{2 r_{1}^2} \left[ -
   \frac{\partial^2}{\partial \chi^2} - \frac{2}{\chi}
   \frac{\partial}{\partial \chi} + \frac{\ell_1 \left( \ell_1 +1
     \right) }{\chi^2} \right]^2 \delta_D \left( \chi - r_1 \right) \, . \quad
 \label{intk6}
 \eea
To solve the integrals~\eqref{intk4}
 and~\eqref{intk6} we have used the identity
 (\ref{identity_bessel}). Hence we obtain \bea b_{\ell_1 \ell_2
   \ell_3} &=& 2 \left( 4\pi \right)^2 \int \frac{dk_2}{k_2}
 \frac{dk_3}{k_3} T_\delta \left( k_2, z_1\right) T_\delta \left( k_3,
   z_1\right) T_\delta \left( k_2, z_2\right) T_\delta \left( k_3,
   z_3\right) P_R \left( k_2 \right) P_R\left( k_3 \right)
 \nonumber \\
 && \qquad \qquad j_{\ell_2 } \left( k_2 r_2 \right) j_{\ell_3 }
 \left( k_3 r_3 \right) j_{\ell_2 } \left( k_2 r_1 \right) j_{\ell_3 }
 \left( k_3 r_1 \right) F_2^{(0)} \left( k_2, k_3 \right)
 \nonumber \\
 &+&2 \left( 4\pi \right)^2 \int d\chi \frac{dk_2}{k_2}
 \frac{dk_3}{k_3} { \frac{\chi^2}{r_1^2}} T_\delta \left( k_2,
   z_1\right) T_\delta \left( k_3, z_1\right) T_\delta \left( k_2,
   z_2\right) T_\delta \left( k_3, z_3\right) P_R \left( k_2 \right)
 P_R\left( k_3 \right)
 \nonumber \\
 && \qquad \qquad j_{\ell_2 } \left( k_2 \chi \right) j_{\ell_3 }
 \left( k_3 \chi \right) j_{\ell_2 } \left( k_2 r_2 \right) j_{\ell_3
 } \left( k_3 r_3 \right) F_2^{(2)} \left( k_2, k_3 \right)
 \nonumber \\
 && \qquad \qquad \left[ - \frac{\partial^2}{\partial \chi^2} -
   \frac{2}{\chi} \frac{\partial}{\partial \chi} + \frac{\ell_1 \left(
       \ell_1 +1 \right) }{\chi^2} \right] \delta_D \left( \chi - r_1
 \right)
 \nonumber \\
 &+&2 \left( 4\pi \right)^2 \int d\chi \frac{dk_2}{k_2}
 \frac{dk_3}{k_3} { \frac{\chi^2}{r_1^2} }T_\delta \left( k_2,
   z_2\right) T_\delta \left( k_3, z_3\right) T_\delta \left( k_2,
   z_1\right) T_\delta \left( k_3, z_1\right) P_R \left( k_2 \right)
 P_R\left( k_3 \right)
 \nonumber \\
 && \qquad \qquad j_{\ell_2 } \left( k_2 \chi \right) j_{\ell_3 }
 \left( k_3 \chi \right) j_{\ell_2 } \left( k_2 r_2 \right) j_{\ell_3
 } \left( k_3 r_3 \right) F_2^{(4)} \left( k_2, k_3 \right)
 \nonumber \\
 && \qquad \qquad \left[ - \frac{\partial^2}{\partial \chi ^2} -
   \frac{2}{\chi} \frac{\partial}{\partial \chi} + \frac{\ell_1 \left(
       \ell_1 +1 \right) }{\chi^2} \right]^2 \delta_D \left( \chi -
   r_1 \right) \nonumber
 \\
 &=&2 \left( 4\pi \right)^2 \int \frac{dk_2}{k_2} \frac{dk_3}{k_3}
 \mathcal{P}_R \left( k_2 \right) \mathcal{P}_R\left( k_3 \right)
 T_\delta \left( k_2, z_1\right) T_\delta \left( k_3, z_1\right)
 T_\delta \left( k_2, z_2\right) T_\delta \left( k_3, z_3\right)
 \nonumber \\
 && \qquad \qquad j_{\ell_2 } \left( k_2 r_2 \right) j_{\ell_3 }
 \left( k_3 r_3 \right) j_{\ell_2 } \left( k_2 r_1 \right) j_{\ell_3 }
 \left( k_3 r_1 \right) F_2^{(0)} \left( k_2, k_3 \right)
 \nonumber \\
 &+&2 \left( 4\pi \right)^2 \int \frac{dk_2}{k_2} \frac{dk_3}{k_3}
 \mathcal{P}_R \left( k_2 \right) \mathcal{P}_R\left( k_3 \right)
 T_\delta \left( k_2, z_1\right) T_\delta \left( k_3, z_1\right)
 T_\delta \left( k_2, z_2\right) T_\delta \left( k_3, z_3\right)
 \nonumber \\
 && \qquad \qquad \frac{1}{ r_1^2} D_{\ell_1} \left[ j_{\ell_2 }
   \left( k_2 \chi \right) j_{\ell_3 } \left( k_3 \chi \right) {
     \chi^2} \right]_{\chi=r_1} j_{\ell_2 } \left( k_2 r_2 \right)
 j_{\ell_3 } \left( k_3 r_3 \right) F_2^{(2)} \left( k_2, k_3 \right)
 \nonumber \\
 &+&2 \left( 4\pi \right)^2 \int \frac{dk_2}{k_2} \frac{dk_3}{k_3}
 \mathcal{P}_R \left( k_2 \right) \mathcal{P}_R\left( k_3 \right)
 T_\delta \left( k_2, z_1\right) T_\delta \left( k_3, z_1\right)
 T_\delta \left( k_2, z_2\right) T_\delta \left( k_3, z_3\right)
 \nonumber \\
 && \qquad \qquad \frac{1}{ r_1^2} D_{\ell_1}^2 \left[ j_{\ell_2 }
   \left( k_2 \chi \right) j_{\ell_3 } \left( k_3 \chi \right) {
     \chi^2} \right]_{\chi=r_1} j_{\ell_2 } \left( k_2 r_2 \right)
 j_{\ell_3 } \left( k_3 r_3 \right) F_2^{(4)} \left( k_2, k_3 \right)
 \eea This reduces the integral from a 4-dimensional
 integral~\eqref{7} to a product of 1-dimensional integrals, since the
 kernels $F^{(i)}_2\left( k_2 , k_3 \right)$ with $i=0,2,4$ are
 separable in $k_2$ and $k_3$. This result agrees with the analytical
 derivation of~\cite{DiDio:2015bua}.
To introduce the linear redshift space distortions we use the
substitution given  in eq.~\eqref{include_RSD}.

%%%%%%%%%%%%%%%%%%%%%%%%%%%%%%%%%%%%
\section{Generalized spectra geometrical factors}
\label{sec:gener-spectra-geom}

We define the geometrical factors used to compute the density
bispectrum dipole, eq.~(\ref{eq:b_dens_dip}), and quadrupole,
eq.~(\ref{eq:b_dens_quad}). For details see appendix A of
\cite{DiDio:2015bua}.

The factor $g_{\ell_1\ell_2\ell_3}$ relates the reduced bispectrum
$b_{\ell_1\ell_2\ell_3}$ to the angle-averaged one. It is given by
\be
g_{\ell_1\ell_2\ell_3}=\sqrt{\frac{(2\ell_1+1)(2\ell_2+1)(2\ell_3+1)}{4\pi}} \tj{\ell_1}{\ell_2}{\ell_3}{0}{0}{0}\,.
\ee
One can show that
\bea
\label{eq:gb_red}
g_{\ell_1\ell_2\ell_3} \, b_{\ell_1\ell_2\ell_3} &\equiv& \sqrt{\frac{(2\ell_1+1)(2\ell_2+1)(2\ell_3+1)}{4\pi}} \tj{\ell_1}{\ell_2}{\ell_3}{0}{0}{0} b_{\ell_1\ell_2\ell_3}
\nonumber \\
&=& \sum_{m_1m_2m_3}  \tj{\ell_1}{\ell_2}{\ell_3}{m_1}{m_2}{m_3} B_{\ell_1\ell_2\ell_3}^{m_1m_2m_3} \;,
\eea
Note that vanishing $g_{\ell_1\ell_2\ell_3}$ values correspond to
unphysical multipole combinations, for which the bispectrum is undefined.

The factor
$Q_{\ell\ell^{\prime }\ell^{\prime \prime }}^{\ell_1\ell_2\ell_3}$ can
be written in terms of the Wigner 6$j$ symbol as
\begin{eqnarray}
\label{eq:Q_def}
Q_{\ell\ell^{\prime }\ell^{\prime \prime }}^{\ell_1\ell_2\ell_3}
=
I_{\ell\ell^{\prime }\ell^{\prime \prime }}^{\ell_1\ell_2\ell_3}\left\{
\begin{array}{ccc}
\ell_1 & \ell_2 & \ell_3 \\
\ell^{\prime} & \ell^{\prime\prime} & \ell
\end{array}
\right\} \left( -1\right) ^{\ell+\ell^{\prime }+\ell^{\prime \prime }} \, ,
\end{eqnarray}
where
$I_{\ell\ell^{\prime }\ell^{\prime \prime }}^{\ell_1\ell_2\ell_3}$
further depends on Wigner 3$j$ symbols:
\be
I_{\ell\ell^{\prime }\ell^{\prime \prime }}^{\ell_1\ell_2\ell_3} \equiv
\sqrt{(4\pi
  )^3(2\ell_1+1)(2\ell_2+1)(2\ell_3+1)}
\left(
\begin{array}{ccc}
\ell & \ell^{\prime \prime } & \ell_1 \\
0 & 0 & 0
\end{array}
\right) \left(
\begin{array}{ccc}
\ell^{\prime } & \ell & \ell_2 \\
0 & 0 & 0
\end{array}
\right) \left(
\begin{array}{ccc}
\ell^{\prime \prime } & \ell^{\prime} & \ell_3 \\
0 & 0 & 0
\end{array}
\right) \;.
\ee
Typically, only a few coefficients of
$Q_{\ell\ell^{\prime }\ell^{\prime \prime }}^{\ell_1\ell_2\ell_3}$
entering in the equations are non-vanishing. They are determined by
the fact that the $6j$ symbol is non-zero only if the triangle
condition is satisfied by all the triplets
\be
(\ell_1,\ell_2,\ell_3),\ (\ell_1,\ell'',\ell),\ (\ell',\ell_2,\ell) \mbox{ and } (\ell',\ell'',\ell_3) \;.
\ee
The $3j$ symbols further require the following sums to be even:
\be
\ell_1+\ell''+\ell,\ \ell'+\ell_2+\ell,\ \ell'+\ell''+\ell_3 \;,
\ee
which also implies
\be
\ell_1+\ell_2+\ell_3 = \text{even} \;.
\ee
To compute Wigner symbols  numerically we use the efficient library
WIGXJPF\footnote{\url{http://fy.chalmers.se/subatom/wigxjpf/}.}
\cite{Johansson:2015cca}.

\section{Cosmic variance for the angular bispectrum}
\label{sec:cv}

In this section we compute the cosmic variance for an arbitrary
redshift-dependent angular bispectrum, generalizing results obtained
in CMB studies \cite{Komatsu:2002db}.  We introduce the bispectrum
estimator (in term of the spherical harmonic expansion coefficients
$a_{\ell m}(z)$) as
\begin{equation}
 { \hat B}_{\ell_1\ell_2\ell_3}(z_1, z_2, z_3) = \sum_{{\textrm{all } m}}
  \tj{\ell_1}{\ell_2}{\ell_3}{m_1}{m_2}{m_3} a_{\ell_1m_1}(z_1)
  a_{\ell_2m_2}(z_2) a_{\ell_3m_3}(z_3)\, .
\end{equation}
This is related to the full bispectrum through
\begin{equation}
{ B_{\ell_1 \ell_2 \ell_3}^{m_1 m_2 m_3} =} \langle a_{\ell_1m_1}(z_1) a_{\ell_2m_2}(z_2) a_{\ell_3m_3}(z_3)
  \rangle = \langle  { \hat B}_{\ell_1\ell_2\ell_3}(z_1, z_2, z_3) \rangle
  \tj{\ell_1}{\ell_2}{\ell_3}{m_1}{m_2}{m_3}\;.
\end{equation}
Analogously we define the power spectrum estimator
\be
\hat c_\ell \left( z_1, z_2 \right) =\left(2 \ell +1 \right)^{-1} \sum_m a_{\ell m}\left( z_1\right) a^*_{\ell m} \left( z_2 \right)
\ee
such that
\begin{equation}
{ c_{\ell}(z_1, z_2)=\langle a_{\ell m}(z_1) a^*_{\ell m}(z_2) \rangle = \langle { \hat c}_{\ell_1}(z_1, z_2) \rangle .}
\end{equation}

Assuming weak non-Gaussianities, the main contribution to the
bispectrum cosmic variance comes from the dominant Gaussian part of
the harmonic expansion coefficients $a_{\ell m}$. In this
approximation
$\langle { \hat B}_{\ell_1\ell_2\ell_3}(z_1, z_2, z_3) \rangle \approx 0$ and
the covariance is given by
\begin{eqnarray}
  && \langle { \hat B}_{\ell_1\ell_2\ell_3}(z_1, z_2, z_3)
     { \hat B}_{\ell_1'\ell'_2\ell'_3}(z'_1, z'_2, z'_3) \rangle =
     \sum_{\textrm{all }mm'} \tj{\ell_1}{\ell_2}{\ell_3}{m_1}{m_2}{m_3}
     \tj{\ell'_1}{\ell'_2}{\ell'_3}{m'_1}{m'_2}{m'_3}\nonumber \\
  &&\quad \langle a_{\ell_1m_1}(z_1) a_{\ell_2m_2}(z_2) a_{\ell_3m_3}(z_3) a^*_{\ell'_1m'_1}(z'_1) a^*_{\ell'_2m'_2}(z'_2) a^*_{\ell'_3m'_3}(z'_3) \rangle
\end{eqnarray}
Wick's theorem gives 15 permutations. However, those permutations
involving terms such as
$\langle a_{\ell_im_i}(z_i) a_{\ell_jm_j}(z_j) \rangle = (-1)^{m_j}
\langle a_{\ell_im_i}(z_i) a^*_{\ell_j -m_j}(z_j) \rangle$ (or,
similarly,
$\langle a^*_{\ell_im_i}(z_i) a^*_{\ell_jm_j}(z_j) \rangle$)
vanish. Recalling that $\tj{\ell_1}{\ell_2}{\ell_3}{m_1}{m_2}{m_3}$ is
zero unless $m_1+m_2+m_3=0$, it is easy to show that such terms give a
contribution to the covariance proportional to
\begin{equation}
  (-1)^m \tj{\ell}{\ell}{\ell'}{m}{-m}{0} =
  \frac{(-1)^\ell}{\sqrt{2\ell+1}} \delta_{\ell'0} \;,
\end{equation}
which vanishes if all multipoles are larger than zero, as in our
case.\footnote{Bispectrum multipoles $\ell \leq 2$ involve non-linear
  terms at the observer, not treatable within cosmological
  perturbation theory.} There are 6 remaining non-vanishing terms from
Wick's theorem (those that only involve contractions of coefficients
with their complex conjugate
$\langle a_{\ell_im_i}(z_i) a^*_{\ell_jm_j}(z_j) \rangle$). Using the
definitions given above and the orthogonality relation
\begin{equation}
  (2\ell+1) \sum_{m_1m_2} \tj{\ell_1}{\ell_2}{\ell}{m_1}{m_2}{m}
  \tj{\ell_1}{\ell_2}{\ell'}{m_1}{m_2}{m'} = \delta_{\ell\ell'} \delta_{mm'}\;,
\end{equation}
together with a more compact notation for power spectra
$c_{\ell}^{ij'} \equiv c_{\ell}(z_i, z'_j)$, the bispectrum covariance
reads
\begin{eqnarray} \label{eq:B_covarince}
  &&\langle
     { \hat B}_{\ell_1\ell_2\ell_3}(z_1, z_2, z_3)
     { \hat B}_{\ell_1'\ell'_2\ell'_3}(z'_1, z'_2, z'_3) \rangle =
     \nonumber \\
  && \quad
     c_{\ell_1}^{11'}   c_{\ell_2}^{22'}
      c_{\ell_3}^{33'}
     \delta_{\ell_1\ell_2\ell_3}^{\ell_1'\ell'_2\ell'_3}
     +
     c_{\ell_1}^{12'}   c_{\ell_2}^{23'}
      c_{\ell_3}^{31'}
     \delta_{\ell_1\ell_2\ell_3}^{\ell_2'\ell'_3\ell'_1}
     +
     c_{\ell_1}^{13'}   c_{\ell_2}^{21'}
      c_{\ell_3}^{32'}
     \delta_{\ell_1\ell_2\ell_3}^{\ell_3'\ell'_1\ell'_2}
     \nonumber \\
  && \quad + (-1)^{\ell_1+\ell_2+\ell_3} \left[
     c_{\ell_1}^{11'}   c_{\ell_2}^{23'}
      c_{\ell_3}^{32'}
     \delta_{\ell_1\ell_2\ell_3}^{\ell_1'\ell'_3\ell'_2}
      +
     c_{\ell_1}^{12'}   c_{\ell_2}^{21'}
      c_{\ell_3}^{33'}
     \delta_{\ell_1\ell_2\ell_3}^{\ell_2'\ell'_1\ell'_3}
  +
     c_{\ell_1}^{13'}   c_{\ell_2}^{22'}
      c_{\ell_3}^{31'}
     \delta_{\ell_1\ell_2\ell_3}^{\ell_3'\ell'_2\ell'_1} \right] \;,
\end{eqnarray}
where
$\delta_{\ell_i\ell_j\ell_k}^{\ell_p\ell_q\ell_r} \equiv
\delta_{\ell_i\ell_p} \delta_{\ell_j\ell_q}
\delta_{\ell_k\ell_r}$. Hence, in the Gaussian approximation the
covariance is diagonal in multipole space. It is useful to consider
the variance for $\ell_1+\ell_2+\ell_3 =$ even:
\begin{eqnarray} \label{eq:B_covarince_sumleven}
  &&\sigma^2_{B_{\ell_1\ell_2\ell_3}}(z_1, z_2, z_3) =\langle
     [{ \hat B}_{\ell_1\ell_2\ell_3}(z_1, z_2, z_3)]^2 \rangle =
     c_{\ell_1}^{11}   c_{\ell_2}^{22}
      c_{\ell_3}^{33}
     + \left[
     c_{\ell_1}^{12}   c_{\ell_2}^{23}
      c_{\ell_3}^{31}
     +
     c_{\ell_1}^{13}   c_{\ell_2}^{21}
      c_{\ell_3}^{32}  \right]
     \delta_{\ell_1\ell_2} \delta_{\ell_2\ell_3}
     \nonumber \\
  && \quad +
     c_{\ell_1}^{11}   c_{\ell_2}^{23}
      c_{\ell_3}^{32}
     \delta_{\ell_2\ell_3}
     +
     c_{\ell_1}^{12}   c_{\ell_2}^{21}
      c_{\ell_3}^{33}
     \delta_{\ell_1\ell_2} +
     c_{\ell_1}^{13}   c_{\ell_2}^{22}
      c_{\ell_3}^{31}
     \delta_{\ell_1\ell_3} \;.
\end{eqnarray}
Hence, the variance strongly depends on whether none, two or all the
multipoles are equal.

At equal redshifts eq.~(\ref{eq:B_covarince}) and
eq.~(\ref{eq:B_covarince_sumleven}) reduce to the usual CMB
expressions \cite{Komatsu:2002db}
\begin{eqnarray}
  \langle { \hat B}_{\ell_1\ell_2\ell_3}(z) { \hat B}_{\ell_1'\ell'_2\ell'_3}(z)
  \rangle &=&  c_{\ell_1}(z)   c_{\ell_2}(z)
               c_{\ell_3}(z)  \left[
              \delta_{\ell_1\ell_2\ell_3}^{\ell_1'\ell'_2\ell'_3} +
              \delta_{\ell_1\ell_2\ell_3}^{\ell_2'\ell'_3\ell'_1} +
              \delta_{\ell_1\ell_2\ell_3}^{\ell_3'\ell'_1\ell'_2}
              \right.
              \nonumber \\
          && + \left. (-1)^{\ell_1+\ell_2+\ell_3} \left(
             \delta_{\ell_1\ell_2\ell_3}^{\ell_1'\ell'_3\ell'_2} +
             \delta_{\ell_1\ell_2\ell_3}^{\ell_2'\ell'_1\ell'_3} +
             \delta_{\ell_1\ell_2\ell_3}^{\ell_3'\ell'_2\ell'_1}
             \right) \right] \;,
\end{eqnarray}
and
\begin{equation} \label{eq:same_z_variance}
  \langle
  { \hat B}_{\ell_1\ell_2\ell_3}(z)^2 \rangle =  c_{\ell_1}(z)
   c_{\ell_2}(z)   c_{\ell_3}(z)  \left( 1 + 2
    \delta_{\ell_1\ell_2}\delta_{\ell_2\ell_3} + \delta_{\ell_1\ell_2}
    + \delta_{\ell_2\ell_3} + \delta_{\ell_3\ell_1}\right) \;,
\end{equation}
respectively.

\bibliographystyle{JHEP}
\bibliography{biblio}

\end{document}